\documentclass[pra, twocolumn, aps, superscriptaddress,longbibliography,showpacs,amsmath,amssymb,floatfix]{revtex4-2}
\usepackage[utf8]{inputenc}
\usepackage[english]{babel}
\usepackage[T1]{fontenc}
\usepackage{amsmath}
\usepackage{graphicx}
\usepackage[colorlinks,linkcolor=blue,anchorcolor=blue,citecolor=blue,urlcolor=blue]{hyperref}
\usepackage{mathtools}
\usepackage{algorithm}
\usepackage{algpseudocode}
\usepackage{multirow}
\usepackage{array}
\usepackage{threeparttable}
\usepackage{lipsum}
\usepackage{tabularx}
\usepackage{booktabs}
\usepackage{color}
\usepackage{xcolor}
\usepackage{xspace}
\usepackage{subcaption}

\usepackage{blindtext}

\newcommand{\casql}{CAS Key Laboratory of Quantum Information, School of Physics, University of Science and Technology of China, Hefei, Anhui, 230026, China}

\newcommand{\aihf}{Institute of Artificial Intelligence, Hefei Comprehensive National Science Center, Hefei, Anhui, 230088, China}

\newcommand{\IAT}{Institute of the Advanced Technology, University of Science and Technology of China, Hefei, Anhui, 230088, China}

\newcommand{\GateiSCZ}{\textsf{iSCZ}}
\newcommand{\GateCiSCZ}{\textsf{C-iSCZ}}

\newcommand{\GateiSWAP}{\textsf{iSWAP}}
\newcommand{\GateSWAP}{\textsf{SWAP}\xspace}
\newcommand{\GateCSWAP}{\textsf{CSWAP}\xspace}
\newcommand{\GateCiSWAP}{\textsf{C-iSWAP}}
\newcommand{\GateCCZ}{\textsf{CCZ}}

\newcommand{\GateIntSWAP}{\textsf{Internal-SWAP}}
\newcommand{\GateRouting}{\textsf{Routing}}

\newcommand{\GateZ}{\textsf{Z}}
\newcommand{\GateY}{\textsf{Y}}
\newcommand{\GateX}{\textsf{X}}
\newcommand{\GateI}{\textsf{I}}
\newcommand{\GateCX}{\textsf{CX}\xspace}
\newcommand{\GateCCX}{\textsf{CCX}}
\newcommand{\GateCZ}{\textsf{CZ}}
\newcommand{\GateS}{\textsf{S}}

\newcommand{\GateSdag}{\textsf{S}^{\dagger}}
\newcommand{\GateCSdag}{\textsf{CS}^{\dagger}}
\newcommand{\GateCSdagone}{\textsf{CS$_{1}$}^{\dagger}}
\newcommand{\GateCSdagtwo}{\textsf{CS$_{2}$}^{\dagger}}

\begin{document}

\title{Hardware-Efficient Quantum Random Access Memory Design with a Native Gate Set on Superconducting Platforms}

\author{Yun-Jie Wang}
\affiliation{\IAT}
\affiliation{\casql}

\author{Sheng Zhang}
\affiliation{\casql}

\author{Tai-Ping Sun}
\affiliation{\casql}

\author{Ze-An Zhao}
\affiliation{\casql}

\author{Xiao-Fan Xu}
\affiliation{\casql}

\author{Xi-Ning Zhuang}
\affiliation{\casql}

\author{Huan-Yu Liu}
\affiliation{\casql}

\author{Cheng Xue}
\affiliation{\aihf}

\author{Peng Duan}
\affiliation{\casql}

\author{Yu-Chun Wu}
\affiliation{\IAT}
\affiliation{\casql}
\affiliation{\aihf}

\author{Zhao-Yun Chen}
\thanks{Corresponding Author: \href{mailto:chenzhaoyun@iai.ustc.edu.cn}{chenzhaoyun@iai.ustc.edu.cn}}
\affiliation{\aihf}

\author{Guo-Ping Guo}
\affiliation{\IAT}
\affiliation{\casql}
\affiliation{\aihf}

\begin{abstract}

Quantum Random Access Memory (QRAM) is a critical component for enabling data queries in superposition, which is the cornerstone of quantum algorithms. Among various QRAM architectures, the bucket-brigade model stands out due to its noise resilience. This paper presents a hardware-efficient native gate set $\{\GateiSCZ, \GateCiSCZ\}$ for implementing bucket-brigade QRAM on superconducting platforms. The experimental feasibility of the proposed gate set is demonstrated, showing high fidelity and reduced complexity. By leveraging the complementary control property in QRAM, our approach directly substitutes the conventional $\{\GateSWAP, \GateCSWAP\}$ gates with the new gate set, eliminating decomposition overhead and significantly reducing circuit depth and gate count.

\end{abstract}

\maketitle
\renewcommand{\thefootnote}{\arabic{footnote}}

\section{Introduction\label{sec: Introduction}}
Quantum random access memory (QRAM)~\cite{Nielsen_Chuang_2010} is essential for quantum algorithms, enabling data queries in superposition—a key requirement for many quantum algorithms~\cite{Grover_1996, HHL_2009, Babbush_Spectra_2018, Huang_DataPower_2021, Biamonte_QML_2017}.
Among the various proposed QRAM architectures~\cite{Low2024tradingtgatesdirty, allcock2023constantdepthcircuitsuniformlycontrolled, phalak_approximateqram_2022, QRAM_Archi_Seth_2008, QRAM_Lloyd_2008, FangYu_Robust_2012}, the bucket-brigade QRAM~\cite{QRAM_Archi_Seth_2008, QRAM_Lloyd_2008, FangYu_Robust_2012} stands out due to its noise resilience.
Similar to classical random access memory (RAM), QRAM has quantum routers to determine the path of the signal qubits to perform the addressing~\cite{QuantumInternet_Kimble2008}. In the quantum circuit model, this quantum router is equivalent to two consecutive $\GateCSWAP$ gates with mutually exclusive conditions, forming what is called a complementary control $\GateCSWAP$ pair~\cite{Uyemura2001}.
The noise resilience is achieved through the use of a quantum router that selectively routes quantum information based on the states of address qubits. Theoretically, QRAM can achieve polylogarithmic scaling of infidelity with memory size~\cite{Arunachalam_robustqram_2015, QRAM_Error_Resilience_Hann_2021}, which ignites the hope of realizing a QRAM. 

Despite its theoretical progress, physical QRAM designs are typically left unspecified~\cite{Aaronson_fineprint_2015, Troyer_hype_2023, jaques_qramsurveycritique_2023}.
Recent advancements of QRAM designs in various platforms, including quantum optics~\cite{Yuan_OpticsRouter_2015}, quantum acoustics~\cite{Connor_Phonon_2019}, and photonics~\cite{Chen_SpinPhotonQRAM_2021}, have shown promise for implementing quantum routers. Given the exponential scalability requirements of routers, the ease of fabrication is a crucial factor in selecting the physical platform for QRAM. Superconducting circuits, known for their straightforward fabrication, have demonstrated quantum supremacy~\cite{GoogleScience_QS_2019, Arute2019-ma} and offer a promising path towards fault-tolerant quantum computers. 

Nevertheless, superconducting circuits face challenges in directly realizing $\GateSWAP$ and $\GateCSWAP$ gates using native interactions.
Intuitively, $\textsf{XY}$ interaction~\cite{Wendin_superconductingtutorial_2017, SuperconductingTutorial_Yvonne_2021, SuperconductingTutorial_Oliver_2021} is a typical type of interaction in the superconducting circuit used to implement the swap-like operations, given by:
\begin{equation}
    H^{XY}_{i, j}=\frac{g^{XY}}{2}\left(X_{i} X_{j}+Y_{i} Y_{j}\right),
    \label{eqn: XY Interaction Hamiltonian Main Text}
\end{equation}
where $X$ and $Y$ are Pauli operators, $g^{XY}$ is the coupling strength.
This interaction generates a two-qubit $\GateiSWAP$ gate after an evolution time of $t_{\text{iSWAP}} = \frac{\pi}{2g}$ (see Appendix~\ref{sec: Native Gate}). This gate swaps the population of two qubits while introducing an extra phase compared to a standard $\GateSWAP$. A similar extra phase will also occur in implementing $\GateCSWAP$.


\begin{figure}[ht]
    \centering
    \includegraphics[width=0.9\columnwidth]{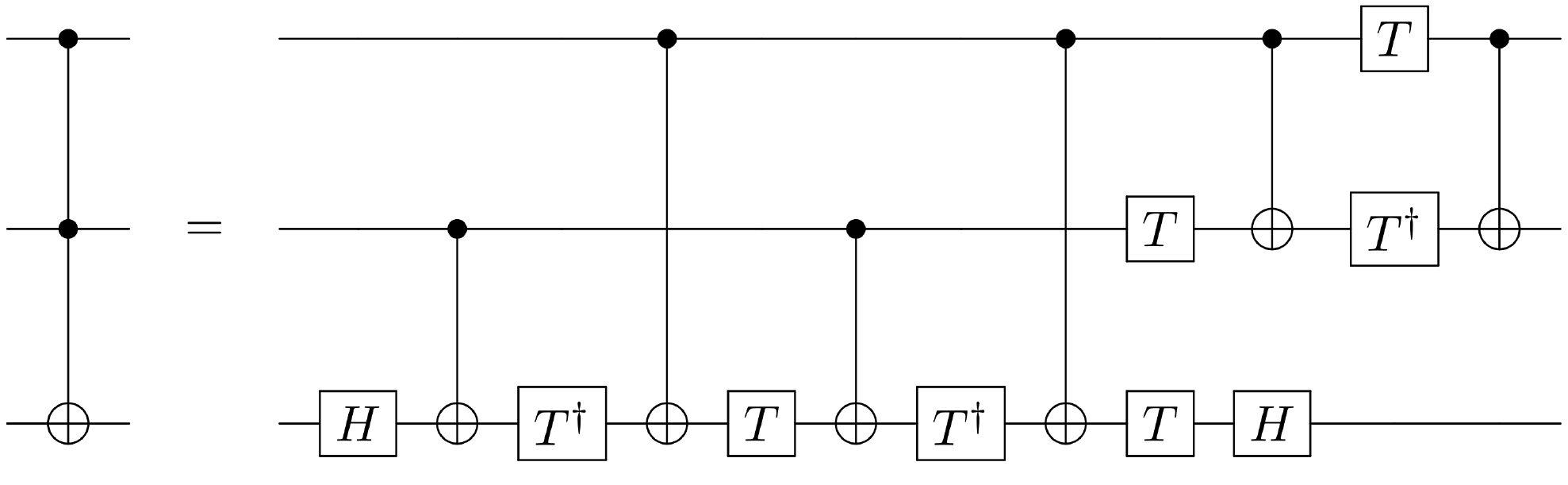}
    \caption{Textbook decomposition of the $\GateCCX$ gate from~\cite{Nielsen_Chuang_2010}. It leads to a circuit with six $\GateCX$ gates and a total depth of 12. The $\GateCSWAP$ is equivalent to one $\GateCCX$ and two $\GateCX$s.}
    \label{fig: Qcircuit_ToffolifromCNOT detailed}
\end{figure}

However, directly using $\GateiSWAP$ and controlled-$\GateiSWAP$ $(\GateCiSWAP)$ as replacements for $\GateSWAP$ and $\GateCSWAP$ in QRAM is not suitable. If no correction is made, the uncorrected phase deviations can accumulate, leading to coherent phase errors in the QRAM query. To overcome this issue, one can decompose the gate into a universal gate set, such as decomposing $\GateCCX$ into $\GateCX$ and single-qubit gates, as shown in Fig.~\ref{fig: Qcircuit_ToffolifromCNOT detailed}. However, this leads to longer gate times, making qubits more susceptible to decoherence~\cite{FredkinIn2Body_Chau_1995, Five_Smolin_1996, Liu-fredkindecompose-2020}.
Recent works use an over-complete gate set with multiple two-qubit native operations, taking advantage of adjustable physical parameters~\cite{Google_fSim_2020, Rigetti_XYGate_2020, Optimized-ZZ}. 
In 3D superconducting cavities, the deterministic $\GateCSWAP$ has been demonstrated by decomposing the three-body operation into three native two-body operations, which reduces gate duration and improves fidelity but requires additional calibration.~\cite{TwoQuantumMemories_Yvonne_2018, Duan_Router_2021, SCQuantumNetwork_Burkhart_2021, Zhou_router_2023, BosonicModes_Yvonne_2019, Weiss_3DCavity_2024}. Furthermore, directly implementing native three-qubit gates, such as $\GateCCZ$, and $\GateCiSWAP$~\cite{Li_CCZ_2019, CCPhase_Filipp_2023, CiSWAP_Zinner_2020} has been demonstrated on superconducting transmon circuits.

Building on recent advancements, we propose a new gate set, $\{\GateiSCZ, \GateCiSCZ\}$, optimized for bucket-brigade QRAM on superconducting platforms. 
Although $\GateiSCZ$ and $\GateCiSCZ$ introduce phase errors compared to the standard QRAM implementation, we address these errors using a phase correction similar to the virtual-Z technique~\cite{VZ_IBM_2017}. Our approach leverages the complementary properties of the quantum router and QRAM structure, resulting in improvements in circuit depth.


Our paper is organized as follows. In Sec.~\ref{sec: QRAM intro}, we provide an overview of the bucket-brigade QRAM, focusing on its underlying principles, circuit architecture, and implementation challenges. In Sec.~\ref{sec: Protocol}, we introduce our proposed gate set, $\{\GateCiSCZ, \GateiSCZ\}$, describing its construction, unique properties, and the improvements it brings to the quantum router. We also discuss the physical feasibility of these gates on tunable transmon qubits, including optimized pulse sequences and simulations using a Kraus model with practical physical parameters. Sec.~\ref{sec: Application} explores optimizations enabled by leveraging complementary control properties, along with an in-depth analysis of the gate set's application in practical QRAM implementations. This section also includes numerical simulations comparing our approach to traditional methods, highlighting gains in efficiency and scalability. Finally, in Sec.~\ref{sec: Conclusion}, we summarize our findings, highlight their impact on improving QRAM scalability and efficiency, and discuss potential experimental realizations as well as further optimizations across various quantum platforms.

\begin{figure*}[ht]
        \centering
        \includegraphics[width=\textwidth]{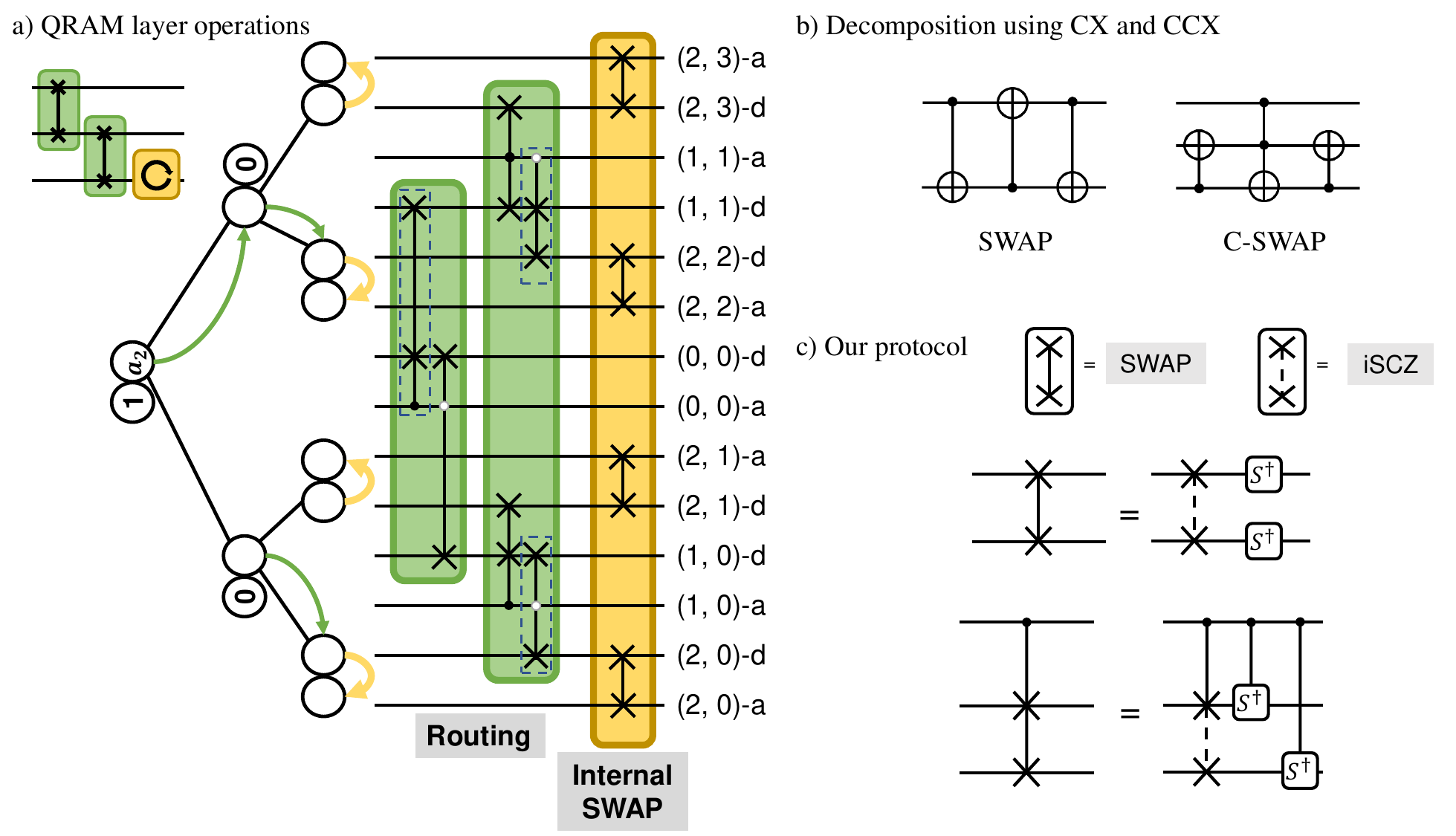}
        \caption{\textbf{(a)} Fundamental operations, $\GateRouting$ and $\GateIntSWAP$, in QRAM (Node: Qubit-Qubit) shown in both the binary tree and quantum circuit models. The figure illustrates the layer-wise $\GateRouting$ and $\GateIntSWAP$ operations (top left) in both models and shows how address information $a_2$ is propagated from the root to the second layer of the ancillary binary tree. Since this is the qubit-qubit version of QRAM, the leftmost (lowest in the figure) path is also activated. Resource requirements for these operations across different layers are summarized in Table.~\ref{tab:summary_qram}. 
        \textbf{(b)} Decomposition of $\GateSWAP$ (left) and $\GateCSWAP$ (right) using $\GateCX$ and $\GateCCX$ gates. The $\GateSWAP$ is realized with 3 $\GateCX$ gates, while $\GateCSWAP$ uses 2 $\GateCX$ gates and 1 $\GateCCX$ gate.
        \textbf{(c)} Schematic representation of $\GateSWAP$ and $\GateCSWAP$ implemented using the proposed gate set ${\GateiSCZ, \GateCiSCZ}$. The $\GateCSWAP$ is decomposed into 1 $\GateCiSCZ$ gate and 2 $\GateCSdag$ gates, while the $\GateSWAP$ requires 1 $\GateiSCZ$ gate and 2 $\GateSdag$ gates.}
        \label{fig:Quantum Circuit Model of QRAM}
\end{figure*}
\begin{figure*}[ht]
    \centering
    \includegraphics[width=\textwidth]{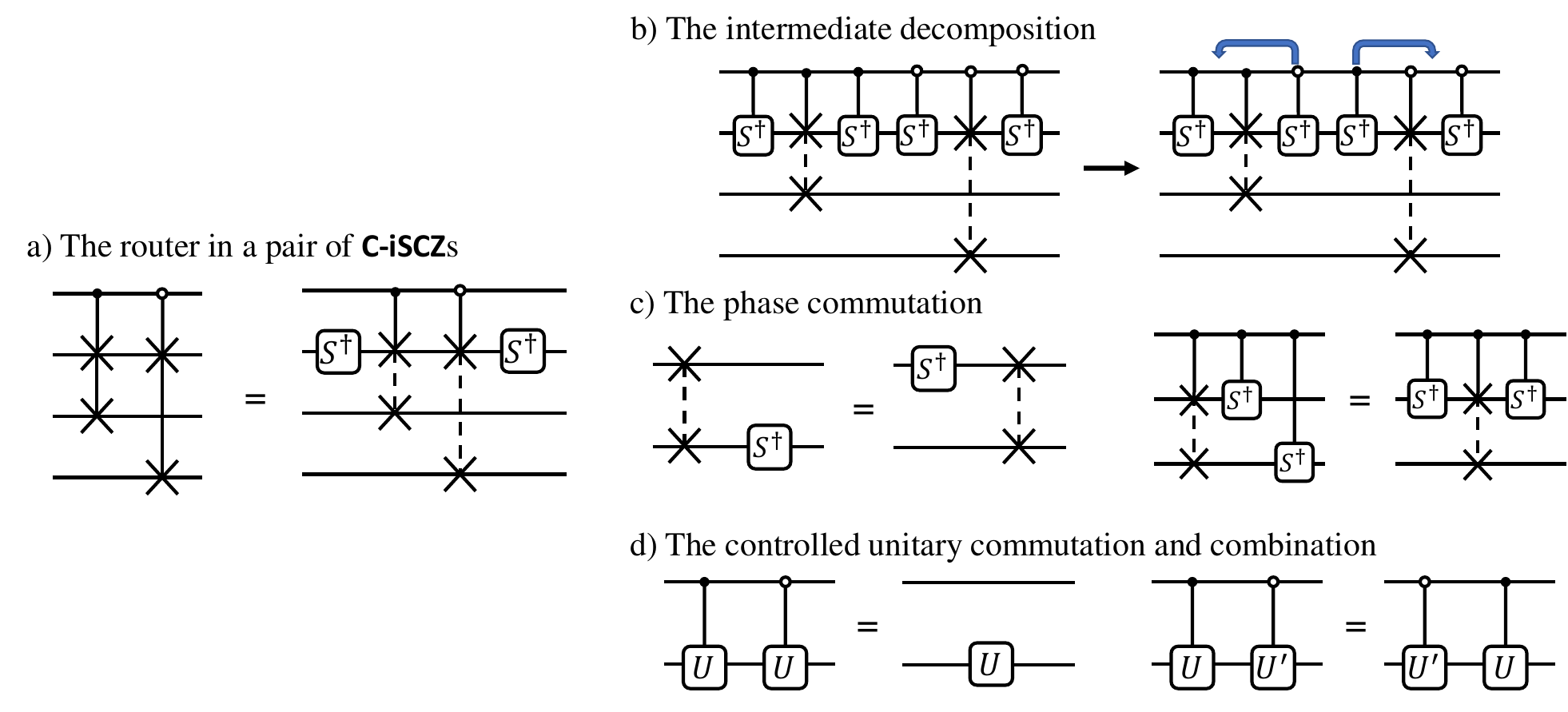}
    \caption{\textbf{(a)} Schematic of the ``complementary control'' $\GateCSWAP$ pairs, representing the router. The optimized implementation uses two $\GateCiSCZ$ gates and two $\GateSdag$ gates, achieving a circuit depth of 4.
    \textbf{(b)} Transformation of the router into the optimized circuit. By applying commutation relations three times and combining properties twice, the total number of multi-qubit operations is reduced from 6 to 2 in (a), eliminating the need for $\GateCSdag$ gates.
    \textbf{(c)} Commutation relations from Eqs.~\eqref{eqn: S and iSCZ Commutation} and ~\eqref{eqn: CSWAP decomposition into C-iSCZ permute}. 
    \textbf{(d)} Commutation and combining properties for general controlled unitary operations. }
    \label{fig: router}
\end{figure*}
\begin{figure*}[ht]
    \centering
    \includegraphics[width=\textwidth]{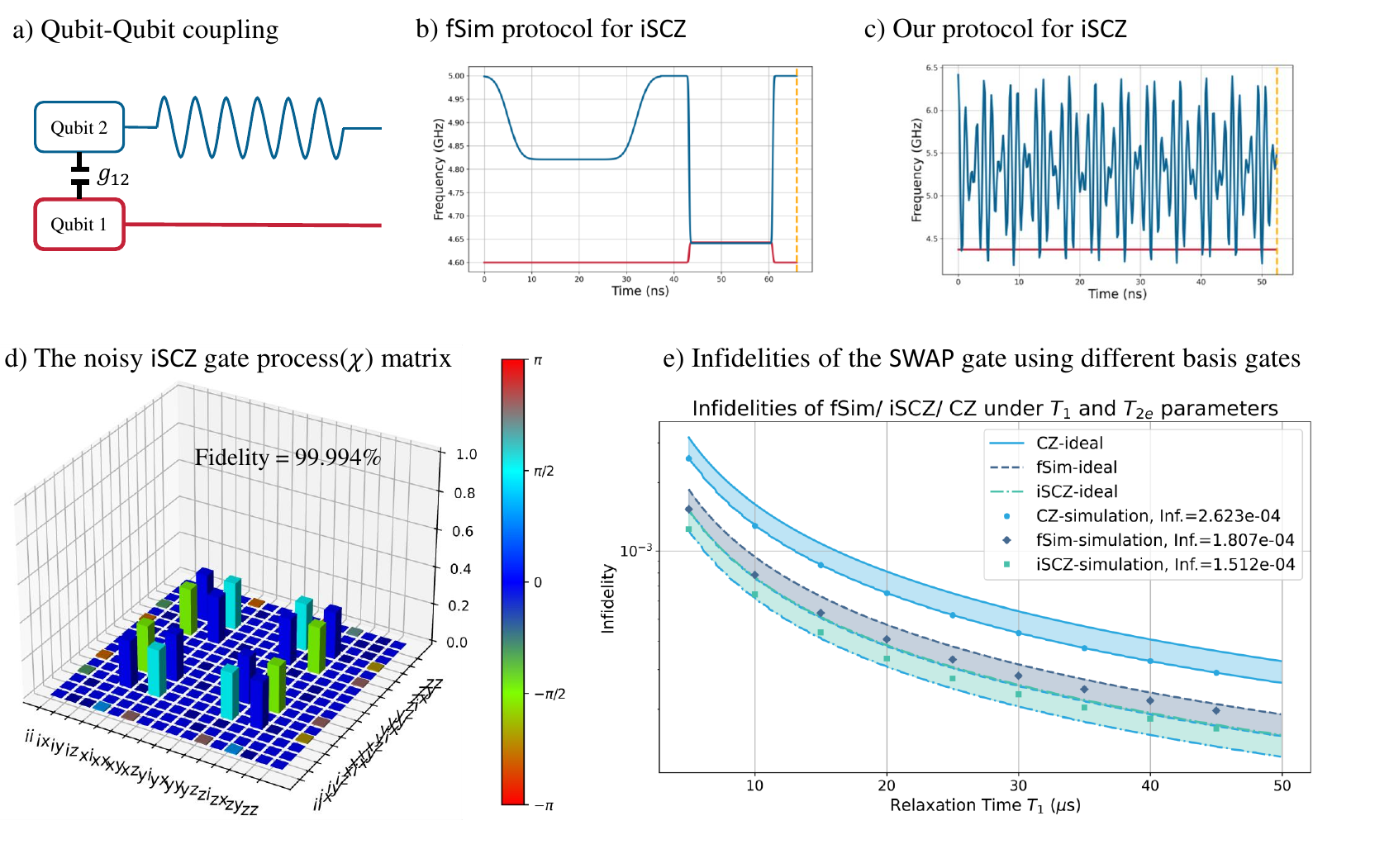}
    \caption{\textbf{(a)} Qubit-qubit coupling to realize $\GateiSCZ$ operations. The red signal represents qubit 1, and the blue sine wave represents qubit 2 in both (b) and (c).
    \textbf{(b)(c)} TDrive pulses for $\GateiSCZ$ using the $\textsf{fSim}$ protocol and our new protocol for both qubits. In the $\textsf{fSim}$ protocol, two sequential pulses are used: the first pulse of 37 ns for $\GateCZ$ and the second pulse of 28 ns for $\GateiSWAP$, resulting in a total duration of 65 ns. In our protocol, a single pulse of 52 ns is used to realized $\GateiSCZ$.
    \textbf{(d)} Quantum process tomography of the $\GateiSCZ$ gate using the mechanism described in Sec.~\ref{subsec: Implementation}. The simulated process matrix $\chi_{\text{sim}}$ is shown in the operator basis ${I \otimes I, I \otimes X, \dots, Z \otimes Z}$. Ignoring single-qubit rotation errors, $\chi_{\text{sim}}$ achieves a fidelity of $\text{Tr}(\chi_{\text{ideal}} \chi_{\text{sim}}) = 99.994\%$. 
    \textbf{(e)} Average fidelity of $\GateCZ$, $\GateiSCZ$, and $\textsf{fSim}$ gates after adding $T_1$ and $T_{2e}$ in Kraus operator simulation. The shaded region represents $T_{2e}$ values ranging from $T_1$ to $2T_1$. The dots correspond to the results using the $\chi$ matrix of noisy gates from our simulation framework with $T_{2e} = 2T_1$.}
    \label{fig: Pulse and Transmon Model}
\end{figure*}


\section{Preliminaries\label{sec: QRAM intro}}
    
    An $(n,k)$-QRAM, which retrieves classical data $d_i$ from address $i$ with a word length of $k$ from one of $N=2^n$ addresses, can be represented by the following unitary operation:

    \begin{equation}
        U_{\mathrm{QRAM}}|i\rangle_A|z\rangle_D=|i\rangle_A\left|z \oplus d_i\right\rangle_D, \quad i \in \left[0,2^n-1\right],
    \end{equation}
    where the address register $A$ consists of $n$ qubits, and the data register $D$ consists of $k$ qubits.

    The naive fan-out QRAM described in Ref.~\cite{Nielsen_Chuang_2010} requires all address nodes to remain coherent during the query, making it vulnerable to decoherence. To mitigate this, the bucket-brigade structure~\cite{QRAM_Lloyd_2008} was introduced, improving infidelity to $\mathcal{O}(n^{2}k)$. A more recent protocol~\cite{QRAM_Parallel_Chen_2023} further improves this to $\mathcal{O}\left((n+k)n\right)$, using the ``load-once'' method for better scaling with word length.

    The bucket-brigade architecture uses an ancillary binary tree to perform the query. Each node contains two qudits: one for the address and one for the data. The address registers can be either three-level ($|W\rangle$, $|L\rangle$, $|R\rangle$) or two-level ($|0\rangle$, $|1\rangle$). 
    
    This structure relies on two key operations, $\GateSWAP$ and $\GateCSWAP$, to transfer data. In the qutrit version, $\GateSWAP$ operations are conditioned on the parent node’s address register, while $\GateRouting$ exchanges information between parent and child nodes. These operations are equivalent to layer-wise $\GateSWAP$ and $\GateCSWAP$ gates, summarized in Fig.~\ref{fig:Quantum Circuit Model of QRAM} and Table~\ref{tab:summary_qram}. Notably, these operations always appear in complementary pairs with mutually exclusive control conditions, a feature termed ``complementary control'' $\GateCSWAP$ pairs, which is crucial for our optimization protocol.

    However, current gate implementations introduce significant overhead. $\GateSWAP$ uses 3 $\GateCX$ gates, and $\GateCSWAP$ requires 2 $\GateCX$ gates plus a Toffoli gate, which itself decomposes into 6 $\GateCX$ gates. This results in deeper circuits and higher error rates, as shown in Fig.~\ref{fig:Quantum Circuit Model of QRAM}(b). Optimizing these operations is essential for reducing errors and improving scalability in QRAM implementations.

    \begin{table}[h]
\caption{Summary of resources required by different layer-wise operations.}
\vspace{0.1cm}
\centering
\begin{threeparttable}
\begin{tabular}{c|c|c}
\toprule
Operation & $\GateCSWAP$s Count & Depth \\ 
\midrule
$\GateRouting(l, l+1)$ & $2^{l+1}$ & $2$    \\ 
\midrule
$\GateIntSWAP(l)$ & $2^{l}$ & 1 (qubit) / 2 (qutrit)  \\
\bottomrule
\end{tabular}
\end{threeparttable}
\label{tab:summary_qram}
\end{table}


\section{Methods\label{sec: Protocol}}
In this section, we first introduce the $\{\GateCiSCZ, \GateiSCZ\}$ gate set as a replacement for the commonly used $\{\GateSWAP, \GateCSWAP\}$ gates in QRAM, as shown in Fig.~\ref{fig:Quantum Circuit Model of QRAM}(a). Next, we describe how to incorporate the ``complementary control'' properties of the quantum router with the proposed gate set. 
Finally, we discuss the feasibility of implementing this gate set on tunable superconducting qubit platforms, completing our analysis.

\subsection{Definitions\label{subsec: Definitions and Properties}}

The $\GateiSCZ$ gate and $\GateS$ are defined as,
    \begin{equation}
        \GateiSCZ = 
        \left[\begin{array}{cccc}
            1 & 0 & 0 & 0 \\
            0 & 0 & i & 0 \\
            0 & i & 0 & 0 \\
            0 & 0 & 0 & -1
        \end{array}\right],
        \quad
        \GateS = 
        \left[\begin{array}{cc}
            1 & 0 \\
            0 & i
        \end{array}\right].
    \label{eqn: Definition of iSCZ}
    \end{equation}
This gate is locally equivalent to the $\GateSWAP$:
\begin{equation}
\GateiSCZ \cdot \left(\GateSdag \otimes \GateSdag \right) = \left(\GateSdag \otimes \GateSdag \right) \cdot \GateiSCZ = \GateSWAP.
    \label{eqn: S and iSCZ Commutation}
\end{equation}

And the $\GateCSWAP$ gate is decomposed into one $\GateCiSCZ$ and two $\GateCSdag$ gates, as follows:
\begin{equation}
    \GateCSWAP = \GateCiSCZ  \left(\GateCSdagone \cdot \GateCSdagtwo\right).
    \label{eqn: CSWAP decomposition into C-iSCZ}
\end{equation}

The native gate set, $\{\GateiSCZ, \GateCiSCZ\}$, reduces circuit depth by introducing a broader range of multi-qubit operations. With this gate set, the $\GateSWAP$ operation can be implemented in a circuit of depth 2 using only one multi-qubit gate, while the $\GateCSWAP$ can be realized with a circuit of depth 3 involving just three multi-qubit gates. This native gate set forms a good start to minimize gate count making it an ideal choice for efficient circuit design. 

\subsection{Router: a complementary control $\GateCSWAP$ pair}
The router, composed of a ``complementary control'' $\GateCSWAP$ pair, contains two $\GateCSWAP$ gates—one conditioned on $|0\rangle$ and the other on $|1\rangle$. Our gate set exploits unique commutation properties to optimize the router. By leveraging these properties, a pair of $\GateCSWAP$s, as shown in Fig.~\ref{fig: router}(a), achieves a depth of 4 with two multi-qubit operations, and the $\GateCSdag$ gate is no longer needed.

The commutation relation between $\GateS$, $\GateSdag$, and $\GateiSCZ$, in Eq.~\eqref{eqn: S and iSCZ Commutation permute}, offers flexibility in circuit design. 
\begin{equation}
    \GateiSCZ \cdot \left(\GateI \otimes \GateSdag \right) =  \left(\GateSdag \otimes \GateI \right) \cdot \GateiSCZ.
    \label{eqn: S and iSCZ Commutation permute}
\end{equation}
Similarly, the same principle applies to $\GateCSdag$ and $\GateCiSCZ$, as expressed in:
\begin{equation}
    \GateCSWAP = \GateCSdag \cdot \GateCiSCZ \cdot \GateCSdag.
\label{eqn: CSWAP decomposition into C-iSCZ permute}
\end{equation}
These relations, summarized in Fig.~\ref{fig: router}, enable the rearrangement of $\GateS$ and $\GateCSdag$ gates before or after each $\GateiSCZ$ and $\GateCiSCZ$ operation.

The ``complementary control'' property of $\GateCSWAP$ pairs offers greater flexibility in gate arrangement. For example, $\GateCSdag$ gates with the same control condition can be combined into a single-qubit $\GateSdag$ gate, while those with different control conditions commute with each other as in Fig.~\ref{fig: router}(d). With all the properties illustrated in Fig.~\ref{fig: router}, the router turns into a ``complementary control'' $\GateCiSCZ$ pair with two single qubit gates. This optimization reduces the number of multi-qubit operations from 6 to 2, achieving the optimal calibration cost for multi-qubit operations—one gate per $\GateCSWAP$ on average.


\subsection{Feasibility of the gate set on superconducting chips\label{subsec: Implementation}}

Both the $\GateiSCZ$ and $\GateCiSCZ$ gates can be implemented on superconducting transmon qubits using native operations. The $\GateiSCZ$ gate has been demonstrated by combining two sequential pulses to realize $\GateiSWAP$ and $\GateCZ$~\cite{Google_fSim_2020}. Here, we propose an alternative method to realize it using a single modulated waveform, eliminating the need to cascade two pulses.

Through fine-tuning parameters like frequency and amplitude, native operations could be described by the $\exp(-iHt)$ and the interaction Hamiltonian $H$ is defined as:
\begin{align}
    H=\sum_i \sum_j g_je^{i(\Bar{\omega_i}-\delta\omega_j)t} 
     {|\phi_j\rangle\langle \varphi_j|}+\mathrm{H.c.},
\end{align}
where $g_j$ is the effective coupling rate and multiple interactions between target energy levels ($|\phi_j\rangle\langle \varphi_j|$) activate when the drive pulse satisfies corresponding resonance conditions $\Bar{\omega_i}=\delta\omega_j$
~\cite{Roth_2017_Analysis_parametric, Didier_2018_Analytical_parametric, Didier_2018_Parametrically_Activated_Entangling, Mundada_2019_suppresion, Hong_parametric_2020, Didier_Parametric-Resonance_2021}. 
$\Bar{\omega_i}$ is the $i$-th frequency component of modulated pulse,
while $\delta\omega_j$ is the frequency difference between $j$-th
energy level pair ($|\phi_j\rangle$ and $|\varphi_j\rangle$).

In a direct qubit-qubit coupling scheme, frequency-division multiplexing with sufficient sampling precision can realize native operations in a single pulse~\cite{Silveri_modulation_2017}. Here, we simulate $\GateiSCZ$ with this promising technique. The $\GateiSCZ$ interaction of a modulated frequency-tunable qubit pair can be described in the interaction picture as:
\begin{align}
H=\sum_i g_{i}[e^{i(\Bar{\omega_i}-\delta)}{|10\rangle\langle01|}
+\sqrt{2}e^{i(\Bar{\omega_i}-\delta+\eta_2)t}{|11\rangle\langle02|} \nonumber \\ 
+\sqrt{2}e^{i(\Bar{\omega_i}-\delta-\eta_1)t}{|20\rangle\langle11|}+\mathrm{H.c.}],
\end{align}
where $g_i$ is the effective coupling rate, $\delta=\omega_{q1}-\omega_{q2}$ is the frequency difference, and $\eta_k$ represents the anharmonicity of the $k$-th qubit. 

The waveform search is done using the differential evolution (DE) and the Nelder Mead (NM) optimization method together~\cite{ControlPulseForm_Machnes_2018, ToffoliOptimization_Zahedinejad_2015, NMOpt_Nelder_1965}. 
During the search, the microwave drive $W$ is parametrized by the total time $t$ and two sinusoidal with fixed frequencies $\omega_{1}, \omega_{2}$ and square pulse envelopes $\Omega_{1}(t), \Omega_{2}(t)$,
\begin{equation}
    W(t, \omega_{1}, \omega_{2}) = \Omega_1(t) \cos{\omega_{1}t} + \Omega_2(t) \cos{\omega_{2}t}.
    \label{eqn: the sum of two waveform in five parameters}
\end{equation}
The fidelity, used as the fitness function, is defined as,
\begin{equation}
    \frac{1}{n(n+1)} 
    \left[
    \mathrm{Tr} \left(M M^{\dagger}\right) 
    + \left|\mathrm{Tr}\left(M M^{\dagger}\right) \right|^2
    \right],
    \label{eqn: experimental quantum operation fidelity}
\end{equation}
and $M = U^{\prime} U^{\dagger}$, $U^{\prime}$ is the result unitary while the $U$ is the ideal operation. A search over the five-parameter space yielded a simulated fidelity of 99.994\% with a duration of 52 ns, as shown in Fig.~\ref{fig: Pulse and Transmon Model}(d).

Following the construction of a $\GateSWAP$ from an $\GateiSCZ$ gate, the $\GateCiSCZ$ operation is a composition of a $\GateCiSWAP$ followed by a $\GateCCZ$. The physical realization of both $\GateCCZ$ and $\GateCiSWAP$ gates has been explored in~\cite{Li_CCZ_2019, NativeCiSWAP_Yu_2023}, achieving fidelities of 93.3\% and 92.36\%, respectively, in 78.5 ns and 59 ns. The duration of $\GateCiSWAP$ gate could be further reduced to 49 ns in Ref.~\cite{Feng_SuperconductingFredkin_2020}. Taking an example of $\GateCZ$ gate time beimg 34 ns in Ref.~\cite{Google_QEC_2023}, the gate decomposition scheme double the operation time compared to our approach.

The $\{\GateiSCZ, \GateCiSCZ\}$ gate set offers a native, efficient alternative to conventional ${\GateSWAP, \GateCSWAP}$ operations in QRAM circuits, leveraging the tunability of superconducting qubits to achieve significant optimization.

\begin{figure*}[ht]
    \centering
    \includegraphics[width=\textwidth]{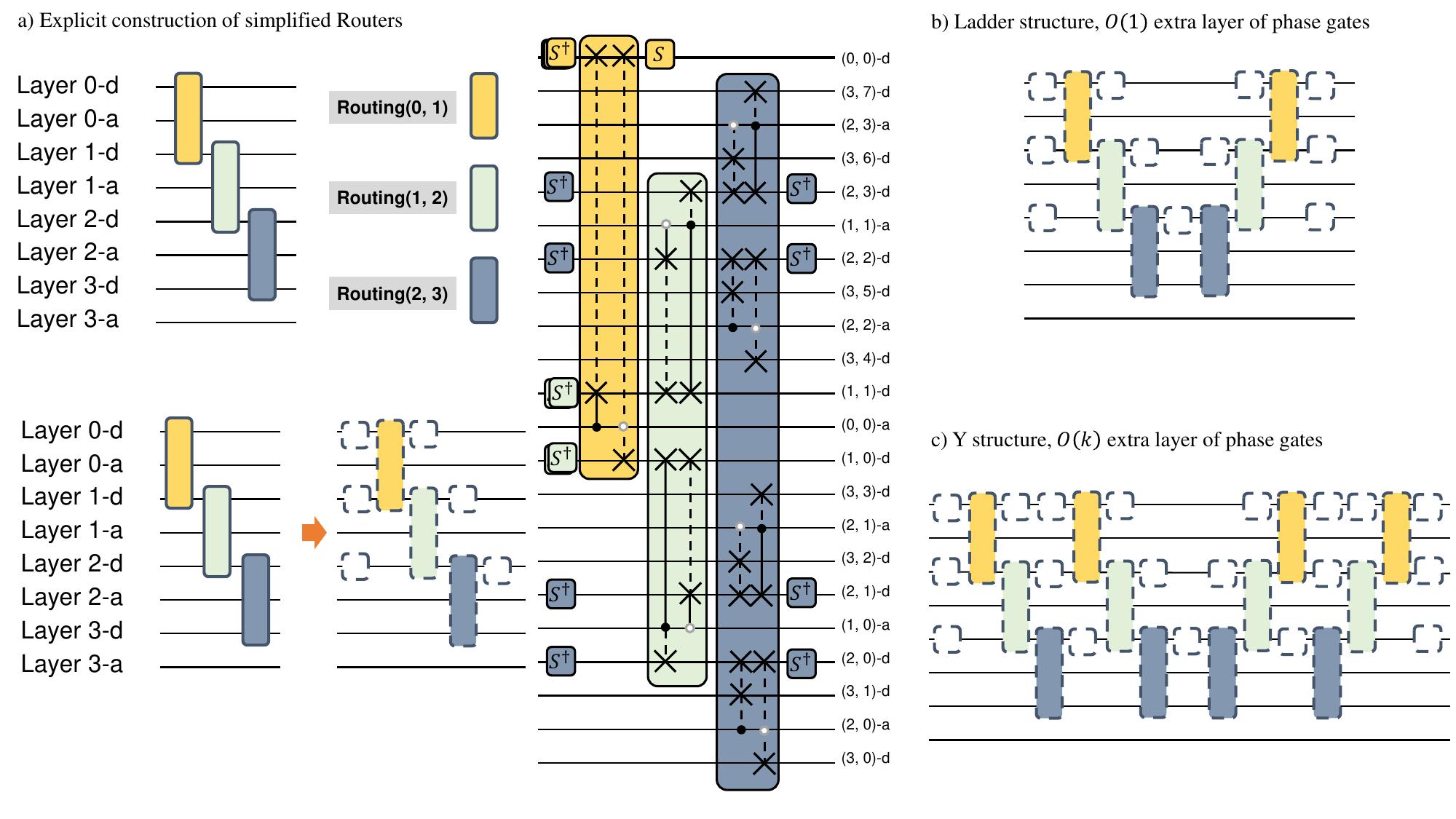}
    \caption{\textbf{(a)} Step-by-step illustration of substituting $\GateRouting$ operations within the same time step. This substitution introduces an extra layer of $\GateSdag$ gates, which can be moved before or after the resulting $\GateCiSCZ$ layer due to the commutation relations in Eq.~\eqref{eqn: S and iSCZ Commutation permute}.
    \textbf{(b)} Ladder structure in a QRAM of depth 3. Due to the commutation relation described in Section~\ref{fig: Virtualizing Phase gates to the start and end of the QRAM circuit}, the extra phase gates can be pushed to the start and end of the circuit, resulting in only a constant number of phase gate layers.
    \textbf{(c)} ``Y'' structure in a QRAM circuit of depth 3, commonly found in $(n,k)$-QRAM when $k > 1$. In this structure, there are $\mathcal{O}(k)$ extra layers of phase gates that are independent of the QRAM depth $n$.}
    \label{fig: direct substitution of C-iSCZ and CS of CSWAP}
\end{figure*}

\section{Full-stack optimization of QRAMs\label{sec: Application}}

Optimizing QRAMs involves more than just improving a single router; it requires a comprehensive protocol for the entire QRAM, which contains multiple routers. The previous section demonstrated how our scheme optimizes quantum routers, but the extra phase gates introduce $\mathcal{O}(n)$ additional layers, significantly increasing execution time.

In this section, we show how to reduce the number of phase gate layers from $\mathcal{O}(n)$ to $\mathcal{O}(1)$, independent of QRAM depth $n$. We also demonstrate that our method simplifies the implementation of a practical QRAM, such as on a 2D grid layout.

    \subsection{``Virtualizing'' extra phase gates\label{subsec: subsec: Optimizing QRAM step 2}}

    As shown in Fig.~\ref{fig: router}(a), two extra phase gates appear before and after the complementary control $\GateCiSCZ$ pair. For QRAMs of depth $n$ (as shown in Fig.~\ref{fig:Quantum Circuit Model of QRAM}), these extra phase gates would increase the circuit depth by $\mathcal{O}(n)$. This is problematic since the original QRAM circuit already has a depth of $\mathcal{O}(n)$. Further reduction is needed to improve performance.

\begin{figure}[ht]
        \centering
        \includegraphics[width=0.9\columnwidth]{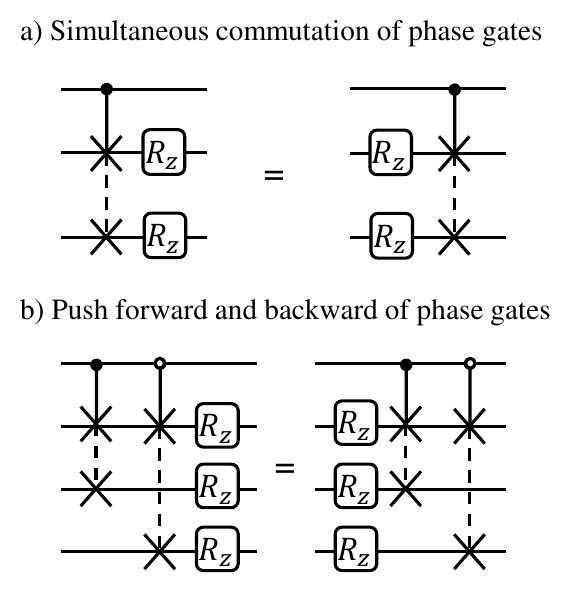}
        \caption{\textbf{(a)} Two $R_z$ gates with the same angle commute with the $\GateiSCZ$. 
        \textbf{(b)} Three $R_z$ gates with the same angle commute through the quantum router without introducing overhead.}
        \label{fig: Virtualizing Phase gates to the start and end of the QRAM circuit}
    \end{figure}

    The key to this reduction lies in the simultaneous commutation of phase gates with $\GateCiSCZ$, as shown in Fig.~\ref{fig: Virtualizing Phase gates to the start and end of the QRAM circuit}(a). The $R_z$ gates, such as $\GateS$, commute with $\GateCiSCZ$. Using this commutation, the phase gates can be pushed to the beginning and end of the circuit, as illustrated in Fig.~\ref{fig: Virtualizing Phase gates to the start and end of the QRAM circuit}(b).

    In QRAM, the ``Y'' and ladder structures (Fig.~\ref{fig: direct substitution of C-iSCZ and CS of CSWAP}(b)(c)) are the main router structures. The phase gates from data registers in layers $l$ and $l-1$ can commute through the $\GateCiSCZ$ gates. This allows them to be moved to the start (before fetching data) and the end (after fetching data) of the circuit as shown in Figs.~\ref{fig: direct substitution of C-iSCZ and CS of CSWAP}.
    
    Moreover, group properties are often employed to simplify circuits~\cite{Clifford_2004_scottdaniel, Bravyi2021cliffordcircuit}. The set $\{\GateS, \GateSdag, \GateZ, \GateI\}$ forms a cyclic group of order four, and these group properties allow phase gates to be shifted within the circuit to further reduce the number of single-qubit operations. For example, the relationships $\GateZ = (\GateSdag)^2$ and $\GateS = (\GateSdag)^3$ provide additional opportunities for gate savings. With these optimizations, the depth overhead from phase gates is reduced to $\mathcal{O}(1)$.

    This method achieves minimal time complexity for bucket-brigade QRAM, particularly in terms of multi-qubit gates. The single-qubit phase gates can be ``virtualized'' to the front and end of the QRAM circuit, ensuring only a constant overhead of single-qubit gates.
    \begin{figure*}[ht]
    \centering
    \includegraphics[width=0.9\textwidth]{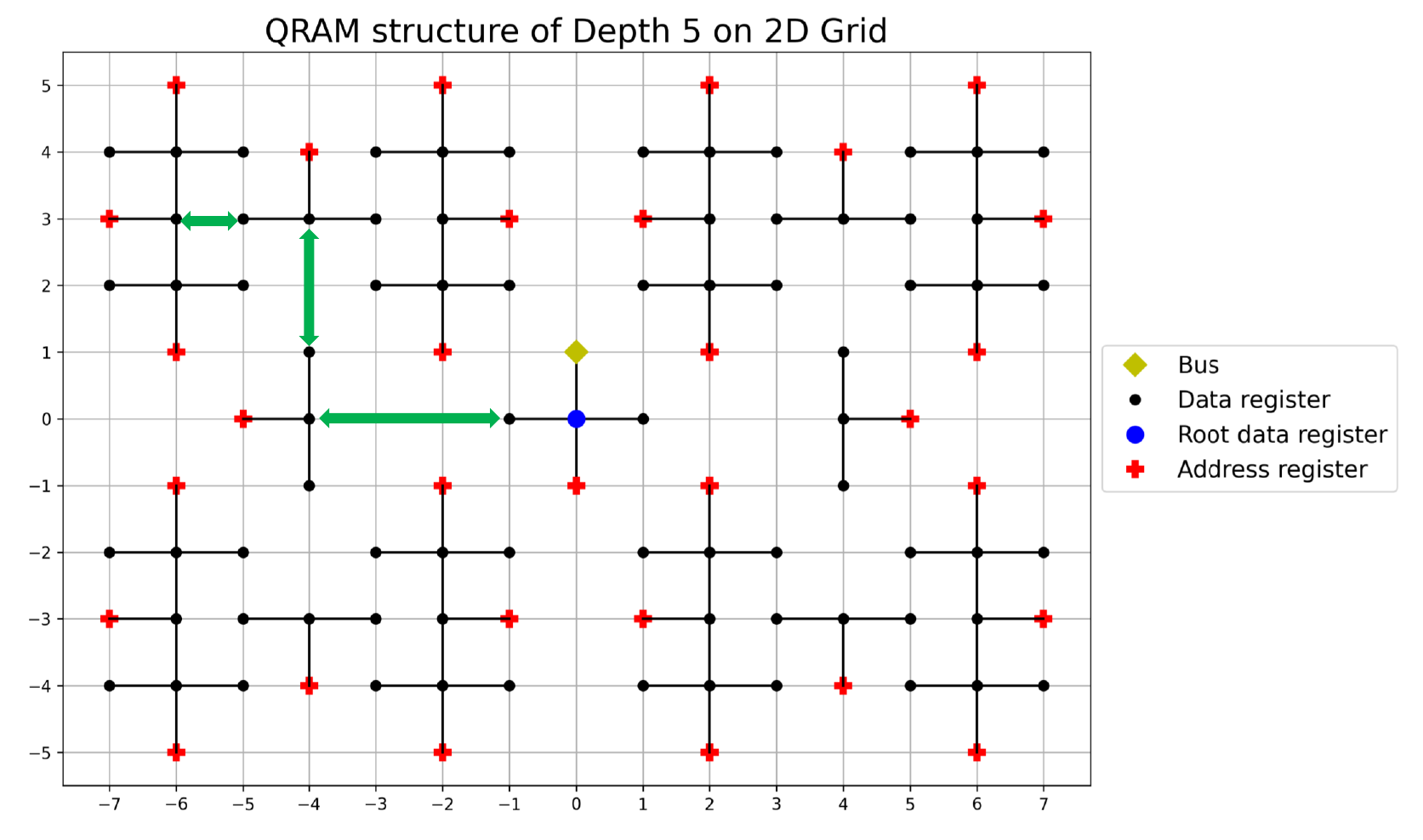}
    \caption{Planar layout of a QRAM circuit with five layers and nearest-neighbor connectivity. The bus (yellow diamond) is located at (0, 1), with the root data register (blue dot) at (0, 0). Red plus signs represent address registers, and black dots represent data registers. Green arrows indicate the distances needed for long-range interactions.}
    \label{fig: A planar layout QRAM-5}
    \end{figure*}
    \subsection{Long-range interaction mediated by $\GateiSCZ$\label{subsec: subsec: Optimizing QRAM step 3}}
        
        Mapping a QRAM circuit onto real hardware presents challenges due to the limited connectivity between qubits in superconducting chips. As QRAM depth increases, the qubit count grows exponentially, complicating circuit layout in a 2D grid. To address this, a modified H-tree layout (Fig.~\ref{fig: A planar layout QRAM-5}) is employed to compactly arrange the circuit. However, this requires long-range interactions, with distances scaling as $\mathcal{O}(\sqrt{N})$, where $N = 2^n$ represents the number of memory cells.

        \begin{figure}[ht]
            \centering
            \includegraphics[width=0.9\columnwidth]{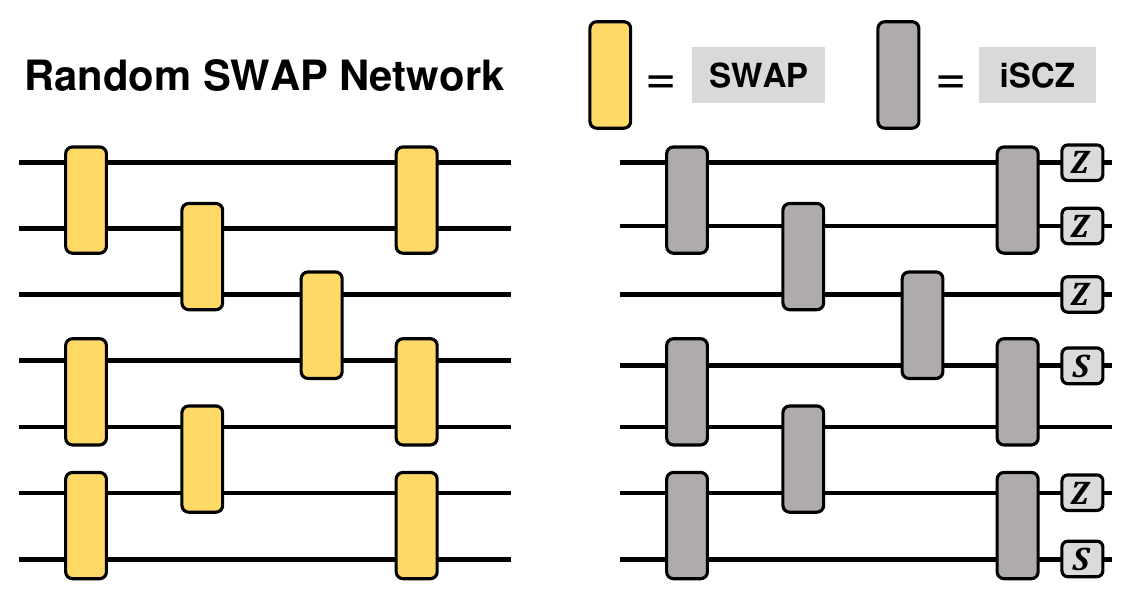}
            \caption{Schematic illustration of the implementation of a 7-qubit SWAP network. The SWAP networks are transpiled using both the $\GateCX$ and $\GateiSCZ$ gates. Based on the Eq.~\eqref{eqn: S and iSCZ Commutation permute}, all phase gates commute through the network to the end of the circuit and form a single layer of phase gates.}
            \label{fig:Numerical Experiments}
        \end{figure}
                
        The most commonly used solution involves the use of permutation circuits, such as SWAP networks, to implement long-range operations while keeping the QRAM passive. Our approach significantly enhances the performance of these permutation circuits by compiling them into shallower circuits compared to those utilizing 3 $\GateCX$ gates for one $\GateSWAP$, thus improving error resilience. Another current approach involves utilizing measurements and conditional classical operations to maintain a constant circuit depth for exchanging information between remote qubits~\cite{Fowler_SurfaceCode_2012, Eddie_SurfaceCodeCompilation_2022, Elisa_longrange_2024}. However, this method requires real-time communication, inevitably transforming the passive QRAM into an active system~\cite{jaques_qramsurveycritique_2023}, thereby negating the asymptotic quantum advantage.
        
        By replacing all $\GateSWAP$ gates with $\GateiSCZ$, we relocate the additional phase gates to the end of the circuit, as illustrated in Fig.~\ref{fig:Numerical Experiments}. This further shortens the depth of the practical QRAMs and avoids the error introduced by decoherence.

        
    \subsection{Numerical Results \label{subsec: Numerical Experiments}}
        We first implement the QRAM circuit and count the two-qubit operations and circuit depth, as shown in Table~\ref{tab: Gate Cost}. The depth increases by 4 as each QRAM layer increases by 1. 

        \begin{table}[h]
\caption{Summary of operation count for realizing $(n, 1)-$QRAM.}
\vspace{0.1cm}
\centering
\begin{threeparttable}
\begin{tabular}{c|c|c|c}
\toprule
Layer$(n)$ & \GateCSWAP & $\GateCiSCZ$ Count$\&$Depth & $\GateCX$ Count$\&$Depth  \\ 
\midrule
2 & 4 & 4/ 6 & 32 / 64   \\ 
\midrule
3 & 12 & 12/ 10 & 96 / 128 \\
\midrule
4 & 28 & 28/ 14 & 224 / 192 \\
\midrule
5 & 60 & 60/ 18 & 480 / 256 \\
\midrule
6 & 124 & 124/ 22 & 992 / 320  \\
\midrule
7 & 252 & 252/ 26 & 2016 / 384 \\
\midrule
8 & 508 & 508/ 30 & 4064 / 448 \\
\bottomrule
\end{tabular}

\end{threeparttable}
\label{tab: Gate Cost}
\end{table}

        \begin{table}[h]
\caption{Summary of operation times}
\vspace{0.1cm}
\centering
\begin{threeparttable}
\begin{tabular}{c|c}
\toprule
Layer & Gate Time  \\ 
\midrule
$\textsf{fSim}$ & 65 ns     \\ 
\midrule
$\GateiSCZ$ & 52 ns  \\
\midrule
$\GateCZ$ & 37 ns  \\
\midrule
$\GateCiSCZ$ & 137 ns  \\
\bottomrule
\end{tabular}

\end{threeparttable}
\label{tab: operation times}
\end{table}

        Gate time analysis in Table~\ref{tab: operation times} shows that realizing $\GateCSWAP$ through $\GateCZ$ decomposition nearly doubles the time compared to using $\GateCiSCZ$, even without accounting for single-qubit gate times. The ratio of gate count reduction (8:1) between direct decomposition and our protocol becomes more significant as the gate count grows exponentially.
        
        Importantly, this improvement is achieved without adding calibration overhead. The number of calibrated gates remains the same for $\GateCiSCZ$ (2) and $\GateiSCZ$ (1) compared to traditional methods. Therefore, our protocol is a hardware-efficient design for QRAM and routers.

        We also tested the SWAP network with qubits arranged linearly, applying $\textsf{SWAP}$ gates exclusively between adjacent qubits. We compared the depth and fidelities of circuits using traditional and optimized decompositions.
        
\begin{figure}[ht]
            \centering
            \includegraphics[width=\columnwidth]{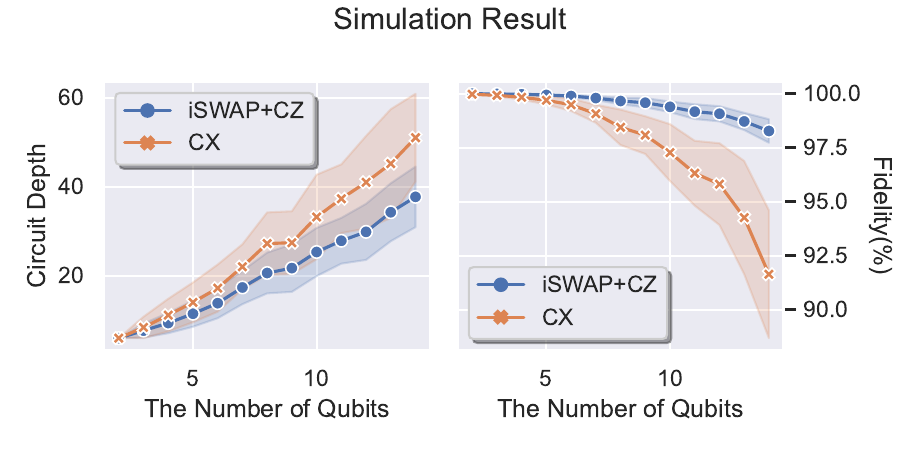}
            \caption{Results of circuit-level simulations of SWAP networks, showing the relationship between fidelity, circuit depth, and the number of qubits. The depolarizing strength is set to $p = 0.02$.}
            \label{fig:Simulation Results}
        \end{figure}

        The fidelity is defined as
        \begin{equation}
            F = \operatorname{Tr}\left[\sqrt{\sqrt{\rho_1} \rho_2 \sqrt{\rho_1}}\right]^2,
            \label{eqn: General Fidelity}
        \end{equation} 
        for both noiseless and noisy circuits, where $\rho_1$ and $\rho_2$ are the density matrices before and after noise is applied. 
        In circuits with a depolarizing strength of 0.02, infidelity decreases threefold, from 9\% to 3\%. As system size increases, our method demonstrates shallower circuit depths and higher fidelities. Despite increased noise, our approach maintains fidelity better than alternative methods as shown in Fig.~\ref{fig:Simulation Results}.
        
        In summary, our protocol leverages the $\{\GateiSCZ, \GateCiSCZ\}$ gate set to achieve substantial reductions in both circuit depth and gate count compared to traditional protocols. The native nature of these gates eliminates the need for decomposing operations into multiple two-qubit gates, thus streamlining the overall design. The inherent commutation properties of our gates further optimize operations like $\GateSWAP$ and $\GateCSWAP$, making our method highly efficient and practical for QRAM implementation.
        
\section{Conclusion and Outlook\label{sec: Conclusion}}

    This paper presents a new framework for implementing bucket-brigade QRAM using the native $\GateiSCZ$ and $\GateCiSCZ$ gate set, replacing conventional $\GateSWAP$ and $\GateCSWAP$ gates. Our approach enables each multi-qubit operation in QRAM to be substituted with a single multi-qubit gate, eliminating the need for gate decompositions and reducing overhead. The protocol is experimentally feasible and has been demonstrated on superconducting qubit platforms, leveraging the natural tunability of transmons to achieve high-fidelity gates. These results confirm the practicality and scalability of our approach for real-world QRAM applications.

    Specifically, our method reduces circuit depth by at least tenfold, e.g. optimizing from 448 to 30 in $n=8$. The multi-qubit gate count is also decreased by eightfold compared to standard decomposition methods, from 4064 to 508 in $n=8$ case. Since the two multi-qubit gates within the set directly and efficiently replace ${\GateSWAP, \GateCSWAP}$ (up to single-qubit operations) in QRAMs, this scheme achieves the optimal number of multi-qubit operations for QRAM.
    
    The $\GateiSCZ$ gate operates in 52 ns with a fidelity exceeding 99.994\%, reducing infidelity by 40\% compared to other single multi-qubit gate methods. The $\GateCiSCZ$ gate, executed in 137.5 ns, effectively replaces $\GateCSWAP$ in QRAM protocols. The relative phase errors caused by these two gates could be systematically tracked and corrected using only single-qubit gates. This gate set of native multi-qubit gates with shorter durations compared to the standard gate set enables high-fidelity QRAMs, particularly in environments where decoherence is the primary source of error.

    We also extend the ``one-gate'' scheme for multi-qubit operations in QRAM by utilizing the complementary control nature of routers. By grouping operations and exploring their commutation properties, we optimize the realization of $\GateCSWAP$ and reduce the need for additional two-qubit gates. While generalizing the ``one-gate'' scheme for all multi-qubit operations remains challenging, our framework shows potential for further optimization.

    We will continue to work on experimentally realizing the efficient and fast $\GateCiSCZ$ gate, expanding the ``one-gate'' scheme to accommodate more qubits, and applying it to a broader range of experimental platforms. Addressing these challenges will advance the scalability of QRAM and quantum computing as a whole.
    
\section{Acknowledgement\label{sec: Acknowledgement}}
    This work has been supported by the National Key Research and Development Program of China (Grant No. 2023YFB4502500), and the National Natural Science Foundation of China (Grant No. 12404564). The numerical experiments are coded in Python using the QISKit and QuTiP library~\cite{Qiskit, QuTiP_2}. We thank Ziwei Cui, and Yongshang Li for the helpful discussions.

\bibliography{mybibliography.bib}

\clearpage
\setcounter{table}{0}
\renewcommand{\thetable}{S\arabic{table}}%
\setcounter{figure}{0}
\renewcommand{\thefigure}{S\arabic{figure}}%
\setcounter{section}{0}
\setcounter{equation}{0}
\renewcommand{\theequation}{S\arabic{equation}}%

\onecolumngrid

\begin{center}
{\large \bf Supplementary Information: Hardware-Efficient Quantum Random Access Memory Design with a Native Gate Set on Superconducting Platforms}\\
\vspace{0.3cm}
\end{center}

\setcounter{page}{1}

\section{Native Gate\label{sec: Native Gate}}
    The Hamiltonian of a quantum computing device is commonly expressed as follows:
    \begin{equation}
        H=\sum_{i=1}^N H_i+\sum_{(i, j)} H_{i, j},
        \label{eqn: General Hamiltonian}
    \end{equation}
    where $N$ is the number of qubits. The first summation term, $H_i$ corresponds to single-qubit operations, while the second term, $H_{i, j}$ describes the interactions between qubits $i$ and $j$.
    
    Depending on the available coupling terms, $H_{i, j}$, different types of two-qubit gates can be implemented as native operations. For Josephson charge qubits coupled via Josephson junctions, the primary interaction Hamiltonian is the XY-interaction:
    \begin{equation}
        H^{XY}_{i, j}=\frac{g^{XY}}{2}\left[X_{i} X_{j}+Y_{i} Y_{j}\right].
        \label{eqn: XY Interaction Hamiltonian}
    \end{equation}
    The $\GateiSWAP$ gate is one of the native operations derived from the XY-interaction. By applying the XY-interaction for a duration of $t = \pi / 2g^{XY}$, the resulting native two-qubit gate is the $\GateiSWAP$ gate:
    \begin{equation}
        \textrm{XY}\left(t\right) =\textrm{exp}\left(-iH^{XY}t\right) = 
        \left[\begin{array}{cccc}
            1 & 0               & 0               & 0 \\
            0 & \cos{g^{XY}t}   & -i\sin{g^{XY}t} & 0 \\
            0 & -i\sin{g^{XY}t} & \cos{g^{XY}t}   & 0 \\
            0 & 0               & 0               & 1
        \end{array}\right] \\.
    \end{equation}

   In addition, the ZZ-interaction serves as the primary interaction Hamiltonian for Josephson charge qubits coupled inductively:
    \begin{equation}
        H_{i, j}=g_{ZZ}\left[Z_{i} Z_{j}\right].
        \label{eqn: ZZ Interaction Hamiltonian}
    \end{equation}
    
    The $\GateCZ$ gate is the native gate derived from the ZZ-interaction. Specifically, $\text{ZZ}\left(\pi / 4g^{ZZ}\right)$ is equivalent to the $\GateCZ$ gate, up to a few single-qubit gates:
    \begin{equation}
        \textrm{ZZ}\left(t\right) =\textrm{exp}\left(-iH^{ZZ}t\right) = 
        e^{-ig^{ZZ}t}
        \left[\begin{array}{cccc}
            1 & 0                            & 0                            & 0 \\
            0 & \exp{\left(i2g^{ZZ}t\right)} & 0                            & 0 \\
            0 & 0                            & \exp{\left(i2g^{ZZ}t\right)} & 0 \\
            0 & 0                            & 0                            & 1
        \end{array}\right] \\.
    \end{equation}

\section{Detailed Derivation of the $\GateiSCZ$ network\label{sec: Detailed Derivation of the iSCZ network}}
There are two main challenges in generalizing the $\GateiSCZ$ decomposition of the SWAP network. 
\begin{figure}[ht]
    \centering
    \includegraphics[scale=0.5]{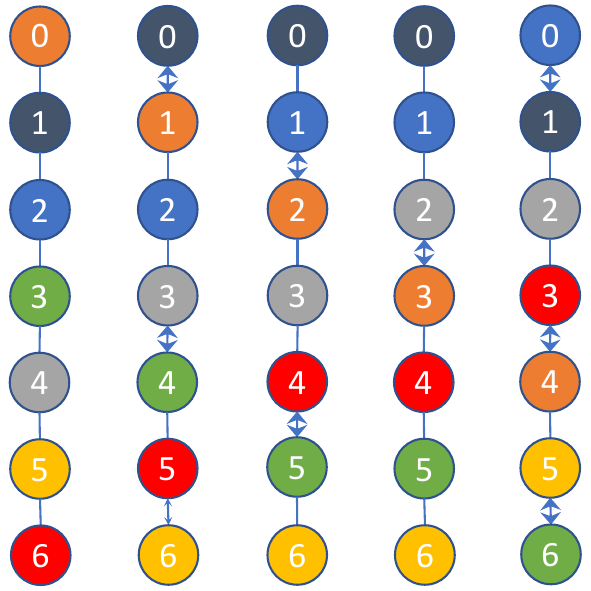}
    \caption{SWAP Network in the Linear Structure}
    \label{fig:SWAP Network in the Linear Structure}
\end{figure}
\begin{figure}[ht]
    \centering
    \includegraphics[width=0.8\columnwidth]{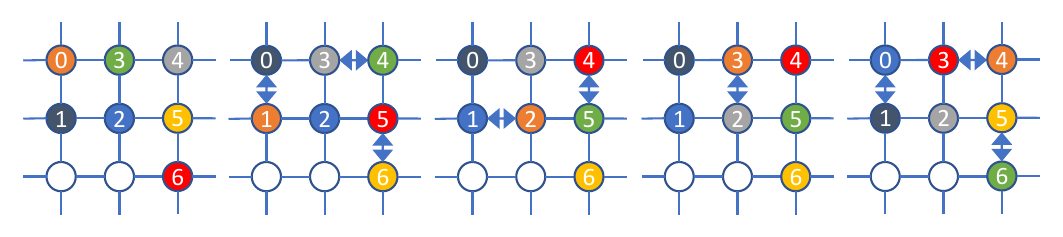}
    \caption{SWAP Network in the Nearest-Neighbor Structure}
    \label{fig:SWAP Network in the NN Structure}
\end{figure}
The first one is about the topology. In our previous analysis, we focused solely on $\GateSWAP$ operations within a linear structure, refer to Fig.~\ref{fig:SWAP Network in the NN Structure}, where we encountered the following limitations: In,
\begin{equation}
\begin{aligned}
    \GateSWAP\left(m, m+1\right) &= \GateiSCZ\left(m, m+1\right) \cdot \GateSdag_{m} \otimes \GateSdag_{m+1} \\
    &= \GateSdag_{m} \otimes \GateSdag_{m+1} \cdot \GateiSCZ\left(m, m+1\right), 
\end{aligned}
\label{eqn: SWAP and iSCZ in linear Structure}
\end{equation}
the indices of two qubits involved in the $\GateSWAP$ operation are not necessarily only differed by 1, such as in the 2D-Grid structure in Fig.~\ref{fig:SWAP Network in the NN Structure}. Therefore, in order to extend the $\GateiSCZ$ formalism to these more general topologies, we need to express $\GateSWAP$, $\GateiSWAP$, and $\GateCZ$ gates in terms of linear combinations of the Pauli operators $\{\GateI, \GateX, \GateY, \GateZ\}$.

\begin{equation}
\begin{aligned}
    \GateSWAP\left(m, l\right) &= \frac{\GateI_{m} \otimes \GateI_{l} + \GateX_{m} \otimes \GateX_{l} + \GateY_{m} \otimes \GateY_{l} + \GateZ_{m} \otimes \GateZ_{l}}{2} \\
    \GateCZ\left(m, l\right) &= \frac{\GateI_{m} \otimes \GateI_{l} + \GateI_{m} \otimes \GateZ_{l} + \GateZ_{m} \otimes \GateI_{l} - \GateZ_{m} \otimes \GateZ_{l}}{2} \\
    \GateiSWAP\left(m, l\right) &= \frac{\GateI_{m} \otimes \GateI_{l} + \GateZ_{m} \otimes \GateZ_{l} + i\left(\GateX_{m} \otimes \GateX_{l} + \GateY_{m} \otimes \GateY_{l}\right)}{2}, 
\end{aligned}
\label{eqn: SWAP and iSCZ general form}
\end{equation}

where $\{ \GateI, \GateX, \GateY, \GateZ\}_i$ represents the corresponding Pauli operators applied to the $i$-th qubit (starting from 0). It can be verified that the relation between $\GateSWAP$ and $\GateiSCZ$, as given in Eq.~\eqref{eqn: SWAP and iSCZ in linear Structure}, still holds for more general topologies when using the Pauli operator representation.

\begin{figure}[ht]
    \centering
    \includegraphics[width=0.4\textwidth]{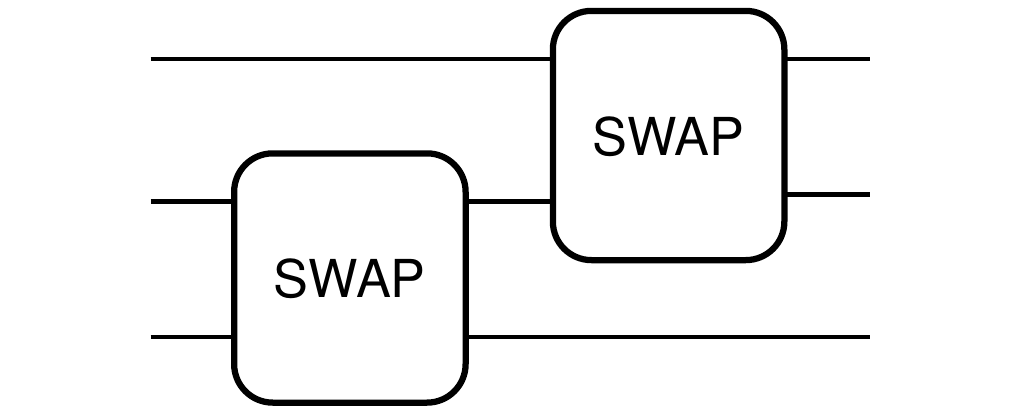}
    \caption{SWAP Circuit 1}
    \label{fig:SWAP Circuit 1}
\end{figure}

\begin{figure}[ht]
  \centering
  \includegraphics[width=0.4\textwidth]{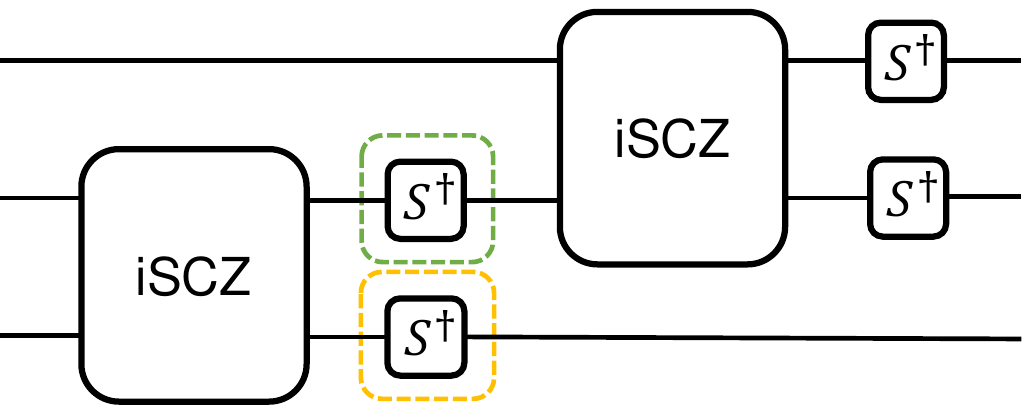}
    \caption{iSCZ Circuit Decomposition}
    \label{fig:iSCZ Circuit Decomposition}
\end{figure}

\begin{figure}[ht]
  \centering
  \includegraphics[width=0.4\textwidth]{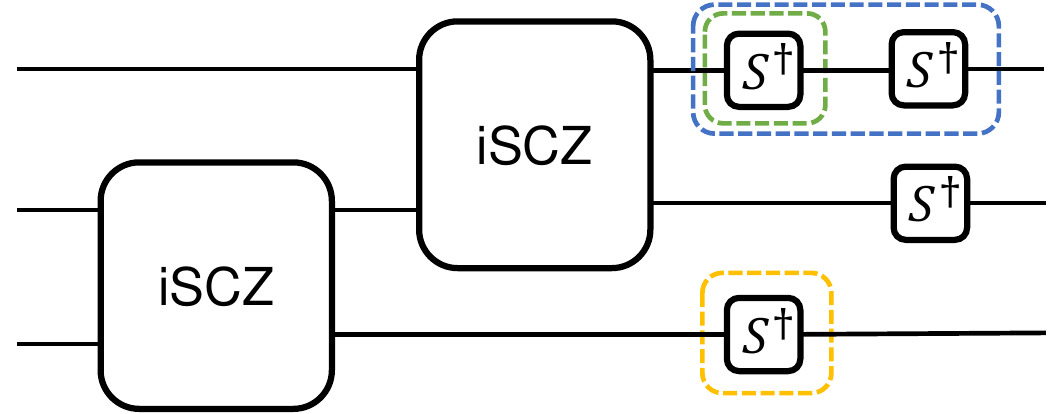}
    \caption{iSCZ Circuit and Phase Decomposition}
    \label{fig:iSCZ Circuit and Phase Decomposition}
\end{figure}

The next problem to solve is how to move the extra phase gates due to the decomposition to the end of the quantum circuit. With the help of the Eq.~\eqref{eqn: S and iSCZ Commutation}, we will have one of the simplest 3-qubit SWAP networks in $\GateiSCZ$ gates in the linear structure demonstrated below(Fig.~\ref{fig:SWAP Circuit 1}), in which we could see how single-qubit phase gates commute throughout the circuit(Fig.~\ref{fig:iSCZ Circuit Decomposition}, Fig.~\ref{fig:iSCZ Circuit and Phase Decomposition}). And Eq.~\eqref{eqn: S and iSCZ Commutation} could also be verified under the Pauli Operator representation(Eq.~\eqref{eqn: SWAP and iSCZ general form}). 

\begin{figure}[ht]
  \centering
  \includegraphics[width=0.4\textwidth]{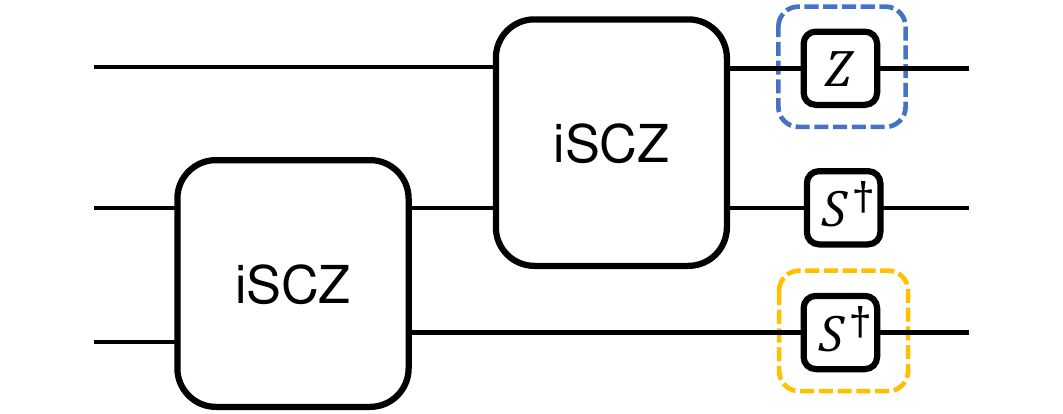}
    \caption{iSCZ Circuit and Phase Decomposition Final}
    \label{fig:iSCZ Circuit and Phase Decomposition Final}
\end{figure}

Assume we have a SWAP network in the form,
\begin{equation}
\begin{aligned}
    &\GateSWAP_{m}\left(A_{m}, B_{m}\right) \cdot \GateSWAP_{m-1}\left(A_{m-1}, B_{m-1}\right) \cdot \cdots \cdot \GateSWAP_{1}\left(A_{1}, B_{1}\right) \\ 
    =& P_{m}P_{m-1} \cdots P_{1} = P_{1 \circ 2 \circ \cdots m},
\end{aligned}
\label{eqn: General SWAP network}
\end{equation}
and $P_{i}$ is defined as the $i$-th permutation(counting from 1) matrix in the SWAP network. The extra phase terms while decomposing the SWAP network could commute through the rest of the network by following the rules below,
\begin{equation}
    \GateSWAP_{g}\left(A_{g}, B_{g}\right) \cdot (\GateSdag_{A_{h}}) = P_{g} \cdot \GateSdag_{A_{h}} = \GateSdag_{P_{g}(A_{h})} \cdot P_{g},
\end{equation}
$\GateSdag_{A_{h}}$ is to apply $\GateSdag$ to the qubit whose index is $A_{h}$ and likewise $\GateSdag_{P_{g}(A_{h})}$ is to apply $\GateSdag$ to the $A_{h}$-th qubit after the permutation $P_{g}$. All phase gates commuted to the end of the circuit could be further simplified due to the periodicity of the S-like phase gate $\left(\GateSdag\right)^2 = \GateZ; \left(\GateSdag\right)^3 = \GateS$. Overall any SWAP networks on any platform could be turned into a network of $\GateiSCZ$ gates and a layer of single-qubit phase gates instead of multiple $\GateCX$s.
\begin{multline}
    \GateSdag_{A_{m}} \GateSdag_{B_{m}} \GateSdag_{P_{m}\left(A_{m-1}\right)} \GateSdag_{P_{m}\left(B_{m-1}\right)} \cdots \GateSdag_{P_{2 \circ \cdots \circ m-1 \circ m}\left(A_{1}\right)} \GateSdag_{P_{2 \circ \cdots \circ m-1 \circ m}\left(B_{1}\right)} \cdot \\
    \GateiSCZ_{m}\left(A_{m}, B_{m}\right) \cdot \GateiSCZ_{m-1}\left(A_{m-1}, B_{m-1}\right) \cdots \GateiSCZ_{1}\left(A_{1}, B_{1}\right).
\label{eqn: Final SWAP network}
\end{multline}

\section{Detailed Derivation of the QRAM Structure\label{sec: Detailed Derivation of the QRAM Structure}}
In this section, I will present the mathematical formalism of the parallel bucket-brigade QRAM and show how it improves the time complexity of data loading. For the $(n,k)$-QRAM, we have states in the form, $|d_{0} d_{1} \cdots d_{k-1} \rangle |l_{0} l_{1} \cdots l_{n-1} \rangle $. $|d_{i}\rangle$ represents one of the data from bus, and $|l_{i}\rangle$ is the data register on the $i$-th layer in the QRAM. Therefore, the entire data loading process for one individual data $d_{i}$ could be written,
\begin{equation}
    |d_{i} \oplus m_{i}\rangle |\textrm{QRAM}\rangle = D^{\dagger}_{i}\left(\prod^{n-2}_{a=0} R^{\uparrow}_{a, a+1}\right) M_{i} \left(\prod^{0}_{b = n-2} R^{\downarrow}_{b, b+1}\right) D_{i} |d_{i}\rangle |\textrm{QRAM}\rangle.
    \label{eqn: QRAM single data}
\end{equation}
First, we will introduce several essential operators and their properties. The operation $M_{i}$ will interact with the data $d_{i}$ at the $(n-1)$-th layer with classical memories to get $d_{i} \oplus m_{i}$. The operation $D_i$ loads the i-th data in the bus to the 0-th layer(root) of the QRAM, and $D^{\dagger}_{i}$ retrieves the data from the data register of the root to the bus.
\begin{itemize}
    \item $D_{i}|d_{0} d_{1} \cdots d_{k-1} \rangle |0\rangle | l_{1} \cdots l_{n-1} \rangle = |d_{0} d_{1} \cdots 0 \cdots d_{k-1} \rangle |d_{i}\rangle |l_{1} \cdots l_{n-1} \rangle$, 
    \item $D^{\dagger}_{i}|d_{0} d_{1} \cdots 0 \cdots d_{k-1} \rangle |d_i\rangle | l_{1} \cdots l_{n-1} \rangle = |d_{0} d_{1} \cdots d_{k-1} \rangle |0\rangle | l_{1} \cdots l_{n-1} \rangle$,
    \item $D_{i}$ does not commute with $D_{j}$ when $i \neq j$.
\end{itemize}
The operation $ R^{\downarrow}_{a, a+1}$ and $R^{\uparrow}_{a, a+1}$ are the same operations that exchange the data on the $a$-th and $(a+1)$-th layer. The arrow indicates how the data $d_{i}$ moves inside the QRAM, and the new data register where the data $d_i$ moves into used to be $| 0 \rangle$.
\begin{itemize}
    \item $R^{\downarrow}_{a, a+1} |d_{i} \rangle_{a} |0 \rangle_{a+1} = |0 \rangle_{a} |d_{i} \rangle_{a+1}$,
    \item $R^{\uparrow}_{a, a+1} |0 \rangle_{a} |d_{i} \rangle_{a+1} = |d_{i} \rangle_{a} |0 \rangle_{a+1}$,
    \item $R^{\uparrow}_{a, a+1}R^{\downarrow}_{a, a+1} = R^{\downarrow}_{a, a+1}R^{\uparrow}_{a, a+1} = \textsc{Routing}\left(a, a+1\right)$,
    \item $R_{a, a+1} R_{b, b+1} = R_{a, a+1} \otimes R_{b, b+1}$, if and only if $|a - b| \geq 2 $.
\end{itemize}
$D_{j}$ and $R_{a, a+1}$, these two types of operations only commutes when $a \geq 1$ because $D_{i}$ is an operation involving the bus and the root. 

Secondly, I will demonstrate the method on optimizing the data loading process and how much it reduces the total time steps in the QRAM structure. It takes $\left(2n+1\right)$ operations for loading each data $d_{i}$. Without any types of optimizations, loading $k$ bits of data requires $\left(2n+1\right)k$ time steps. Instead of loading each data individually in Eq.~\eqref{eqn: Non-Parallel QRAM}, 
\begin{equation}
    \prod^{k-1}_{i=0}  \left[D^{\dagger}_{i}\left(\prod^{n-2}_{a=0} R^{\uparrow}_{a, a+1}\right) M_{i} \left(\prod^{0}_{b = n-2} R^{\downarrow}_{b, b+1}\right) D_{i}\right],
    \label{eqn: Non-Parallel QRAM}
\end{equation}
the new-proposed QRAM structure parallelizes those operations as much as possible so that it could reach better time complexity. By analyzing the commutation relation between $\left(\prod^{0}_{b = n-2} R^{\downarrow}_{b, b+1}\right) D_{i+1}$ and $D^{\dagger}_{i}\left(\prod^{n-2}_{a=0} R^{\uparrow}_{a, a+1}\right)$, we will have,
\begin{equation}
\begin{aligned}
    &\textsc{QRAM}\left(D_{j+1}\right)\textsc{QRAM}\left(D_{j}\right) \\
    &= \prod^{j+1}_{j} \left[D^{\dagger}_{i}\left(\prod^{n-2}_{a=0} R^{\uparrow}_{a, a+1}\right) M_{i} \left(\prod^{0}_{b = n-2} R^{\downarrow}_{b, b+1}\right) D_{i}\right] \\
    &= D^{\dagger}_{j+1}R^{\uparrow}_{0, 1}\left[\left(\prod^{n-2}_{a=1} R^{\uparrow}_{a, a+1} M_{j+1}\right) \otimes \left(D^{\dagger}_{j} \prod^{n-3}_{a=0} R^{\uparrow}_{a, a+1} \right)\right] \cdot \textsc{Routing}\left(n-2, n-1\right) \\
    & \left[\left(\prod^{0}_{b=n-3} R^{\uparrow}_{b, b+1} D_{j+1}\right) \otimes \left(M_{j}\prod^{1}_{n-2}R^{\uparrow}_{b, b+1} \right) \right]R^{\downarrow}_{0, 1} D_{j}\\.
\end{aligned}
\label{eqn: Parallel QRAM NN}
\end{equation}
This relation will minimize the time complexity of loading two bits of data by delaying all operations for loading the second one by two time steps. The time steps are reduced from $2\left(2n+1\right)$ to $\left( 2n+3 \right)$. Besides, the Eq.~\eqref{eqn: Parallel QRAM  General} not only considers the merge of operations for loading adjacent data but also for loading data that is not next to each other.
\begin{equation}
\begin{aligned}
    &\textsc{QRAM}\left(D_{j+k}\right)\textsc{QRAM}\left(D_{j}\right) \\
    &= D^{\dagger}_{j+k}
    \prod^{2k-2}_{m=0}R^{\uparrow}_{m, m+1}\left[\left(\prod^{n-2}_{a=2k-1} R^{\uparrow}_{a, a+1} M_{j+k} \prod^{n-k-2}_{l=n-2}R^{\downarrow}_{l, l+1}\right) \otimes \left(D^{\dagger}_{j} \prod^{n-k-2}_{a=0} R^{\uparrow}_{a, a+1} \right)\right]  \\
    & \cdot \textsc{Routing}\left(n-k-1, n-k\right)\cdot \\
    & \left[\left(\prod^{0}_{b=n-k-2} R^{\uparrow}_{b, b+1} D_{j+k}\right) \otimes \left(\prod^{n-2}_{l=n-k-2}R^{\downarrow}_{l, l+1}M_{j}\prod^{2k-1}_{n-2}R^{\uparrow}_{b, b+1} \right) \right] \prod^{0}_{m=2k-2}R^{\downarrow}_{m, m+1} D_{j}\\.
\end{aligned}
\label{eqn: Parallel QRAM General}
\end{equation}
The previous equation also shows that when $k \geq n$, the loading processes for $d_{j+k}$ and $d_{j}$ have no overlap, which means that all of these operations for loading these two bits of data will be applied sequentially. 

Furthermore, it is clear to see that the time complexity will be $\mathcal{O}\left(n+k\right)$ instead of $\mathcal{O}\left(nk\right)$, which helps a lot when loading a large set of data. When analyzing data loading operations for $k$ consecutive operations from $d_{j+k-1}$ to $d_{j}$, we also find that there will be $\sum^{k-1}_{i=1}i$ pairs of $R^{\uparrow,\downarrow}_{n,n+1}$ operations could be merged together if $k \leq n$ and $\sum^{n-1}_{i=1}i+\left( k-n \right) \left( n-1 \right)$ for a larger k. The analytical evaluation of the new-proposed bucket-brigade QRAM structure has shown how it takes advantage of the power of parallel operations and lays the foundation for optimizations of practical implementations of the QRAM.

\section{QRAM Basics}
Practical implementations of QRAMs, such as those described in~\cite{QRAM_Parallel_Chen_2023} and~\cite{QRAM_Archi_Seth_2008}, typically involves three essential components: the data bus, the data memory, and the ancillary binary tree. The data bus handles incoming and outgoing data, while the data memory contains the data cells to be accessed based on specific addresses. The parallel protocol proposed in~\cite{QRAM_Parallel_Chen_2023} consists of three stages: address setting, data fetching, and uncomputing. During address setting, a specific path is formed from the root to one of its children in the binary tree, guided by address information from the bus. In the data fetching stage, data from the bus traverses this path to the last layer of the tree, interacts with the classical data in the memory cells, and then returns to the data bus along the same path. Uncomputing involves reversing all address setting actions. 
        
        Our approach is highly suitable for optimizing the structure of a QRAM because it essentially functions as a SWAP network. Each node in the ancillary binary tree consists of one qutrit for the address register and one qubit for the data register, denoted as $|N\left(l,m\right)\rangle = |a\left(l,m\right)\rangle \otimes |d\left(l,m\right)\rangle$, where $a\left(l,m\right) \in \left[0, 1, 2\right]$ and $d\left(l,m\right) \in \left[0, 1\right]$. The value 0 in the address register represents a passive or ``wait'' mode for the data qubit, while 1 and 2 signify swapping left or right, respectively. Noted that if qutrits are not accessible, qubits could also be used as address registers. Consequently, for a given layer index $l$, the layer can be expressed as $\bigotimes^{2^l - 1}_{m=0} |N\left(l,m\right)\rangle$. Furthermore, the entire ancillary binary tree, requiring $\mathcal{O}\left(2^n\right)$ qubits, can be expressed as:
        \begin{equation}
            \bigotimes^{n - 1}_{l=0}\bigotimes^{2^l - 1}_{m=0} |N\left(l,m\right)\rangle.
        \label{eqn: QRAM Binary Tree}
        \end{equation}
        And the initial state of the QRAM binary tree is given by $\left(|0\rangle|0\rangle\right) \left(|0\rangle|0\rangle\right)^{\otimes 2} \cdots \left(|0\rangle|0\rangle\right)^{\otimes 2^{n-1}}$. 

        Two operations are repeatedly employed in QRAM implementations~\cite{QRAM_Parallel_Chen_2023} that are essentially $\GateSWAP$s. The first is the $\GateIntSWAP$ operation. The $\GateIntSWAP$ operation alters the state of all address registers within a layer based on the data register of the corresponding node and the address register of its parent node. The \textsf{Routing} operation exchanges the information between the parent node with one of its child nodes based on the status of its address register.
        
        The operation $\GateIntSWAP\left(l\right)$ acts on $2^l$ nodes on the $l$-th layer based on the address register of their parent node in the $\left(l-1\right)$-th layer. In the quantum circuit model, this operation requires $2^l$ $\textsf{controlled-3-2-SWAP}$ gates. For instance, nodes $\left(l, 2n\right)$ and $\left(l, 2n+1\right)$ are modified according to the address register of $\left(l-1, n\right)$. Explicitly, the operation can be written as:
        \begin{equation}
        \begin{aligned}
            &|0 \rangle\langle 0| \otimes \textsc{I}^{\otimes 2} + |1 \rangle\langle 1| \otimes\textsf{3-2-SWAP}\left[a\left(l, 2n\right), d\left(l, 2n\right)\right] \\
            &+|2\rangle\langle 2| \otimes\textsf{3-2-SWAP}\left[a\left(l, 2n+1\right), d\left(l, 2n+1\right)\right],
        \end{aligned}
        \label{eqn: Internal SWAP-Explicit-Single}
        \end{equation}
        where the first qutrit corresponds to the address register of node $\left(l-1, n\right)$ and the $\textsf{3-2-SWAP}$ operation is defined specifically for qutrit-qubit nodes, which accomplishes transitions $|0\rangle_{a} |0\rangle_{d} \Leftrightarrow |1\rangle_{a} |0\rangle_{d}$ and $|0\rangle_{a} |1\rangle_{d} \Leftrightarrow |2\rangle_{a} |0\rangle_{d}$.

        The operation $\textsf{Routing}\left(l, l+1\right)$ acts on $2^l$ nodes on the $l$-th layer to decide which node in the next layer to land specifically the data on its data register based on its address register. In the quantum circuit model, the routing operation $\textsf{Routing}\left(l, l+1\right)$ could be decomposed into $2^l$ pairs of $\GateSWAP$ and $\GateCSWAP$ operations. And each pair has a form like, 
        \begin{equation}
            \begin{aligned}
                &|0 \rangle\langle 0| \otimes \textsc{I}^{\otimes 2} + |1 \rangle\langle 1| \otimes \GateSWAP\left[d\left(l, j\right), d\left(l+1, 2j\right)\right] \\
                &+|2\rangle\langle 2| \otimes\GateSWAP\left[d\left(l, j\right), d\left(l+1, 2j+1\right)\right].
            \end{aligned}
            \label{eqn: Routing-QRAM-Single}
        \end{equation}

            \subsubsection{Layer Representation}
    Due to the unique design of QRAM, there are many operations that could be implemented in parallel. Therefore, we can simplify the expressions of two operations at the layer level. The full description of layer $l$, 
    \begin{equation}
        \bigotimes^{2^l - 1}_{m=0} |a\left(l,m\right)\rangle|d\left(l,m\right)\rangle, 
        \label{eqn: layer description}
    \end{equation} 
    where $a\left(l,m^\prime\right) \neq 0$. The net effect of the operation $\GateIntSWAP\left(l-1, l\right)$ can be written as:
    \begin{equation}
    \begin{aligned}
        &\GateIntSWAP\left(l-1, l\right) \\
        &= \textsf{3-2-SWAP}\left[a\left(l, 2n^{\prime}+a\left(l-1, n^{\prime}\right)-1\right), d\left(l, 2n^{\prime}+a\left(l-1, n^{\prime}\right)-1\right)\right].
    \end{aligned}
    \label{eqn: Internal SWAP-Implicit-General}
    \end{equation}
    Here, the $\textsf{3-2-SWAP}$ operation is defined specifically for qutrit-qubit nodes and accomplishes the transitions $|0\rangle_{a} |0\rangle_{d} \Leftrightarrow |1\rangle_{a} |0\rangle_{d}$ and $|0\rangle_{a} |1\rangle_{d} \Leftrightarrow |2\rangle_{a} |0\rangle_{d}$.
    
    Similarly, the $\textsf{Routing}\left(l, l+1\right)$ could be simplified as, 
    \begin{equation}
        \textsf{Routing}\left(l, l+1\right) = \GateSWAP\left[d\left(l, j^{\prime}\right), d\left(l+1, 2j^{\prime}+a\left(l, j^{\prime}\right)+1\right)\right],
    \label{eqn: Internal SWAP-Implicit-General}
    \end{equation}
    where the node $\left(l, j^{\prime}\right)$ has the only non-zero address register in $l$-th layer. 
    If the address register of the node in the QRAM is qubit instead of qutrit, there will be a minor change in the $\GateIntSWAP$,
    \begin{equation}
        \GateIntSWAP\left(l-1, l\right)|a_{l-1}\rangle|d_{l-1}\rangle |a_{l}\rangle|d_{l}\rangle = |a_{l-1}\rangle|d_{l-1}\rangle |d_{l}\rangle|a_{l}\rangle,
        \label{eqn: InternalSWAP-Qubit-Single}
    \end{equation}
    while the $\textsf{Routing}$ operation remains the same, as in Fig.~\ref{fig:Quantum Circuit Model of QRAM}.
    
\section{Detailed Gate Complexity Analysis\label{sec: Detailed Gate Complexity Analysis}}
In this section, I will summarize the results regarding the gate complexity in the $(n,k)$-QRAM implementations in the qubit-qubit scheme. The gate complexity analysis focuses on three main phases: the address setting, the uncomputing, and the data fetching.

For address setting and uncomputing, a total of $2n$ $\GateIntSWAP$ and $n\left(n-1\right)$ $\textsf{Routing}$ operations are required. During the data fetching process, the number of $\textsf{Routing}$ operations involved can be determined based on the values of $n$ and $k$. Specifically, there are $2(n-1)k - f(n,k)$ $\textsf{Routing}$ operations required, where $f(n,k)$ is defined as:
    \begin{equation}
    f\left(n,k\right) = 
        \begin{cases}
            k\left(k-1\right) / 2 & n \geq k\\
            n\left(n-1\right) / 2 + \left(k-n\right)\left(n-1\right)& n < k
        \end{cases}.  
    \end{equation}
Alternatively, the expression can be simplified as $2(n-1)k - \sum^{k-1}_{i=1}\min \left(i, n\right)$.

However, it is essential to emphasize that the costs associated with $\GateIntSWAP$ and $\textsf{Routing}$ operations can vary across different layers. Specifically, the unidirectional $\textsf{Routing}\left(l, l+1\right)$ entails $2^{l}$ pairs of $\GateCSWAP$ and $\GateSWAP$ operations, whereas bidirectional one requires $2^{l}$ pairs of two $\GateCSWAP$ operations. Additionally, $\GateIntSWAP\left(l-1, l\right)$ necessitates $2^{l-1}$ pairs of $\GateCSWAP$ and $\GateSWAP$ operations. Therefore, in order to accurately assess the gate complexity, it is critical to meticulously count the exact number of $\textsf{Routing}$ and $\GateIntSWAP$ operations performed at each layer. This careful consideration allows for a precise evaluation of the overall gate complexity.

The first operation to analyze is the $\GateIntSWAP$ during the address setting and uncomputing phase. The total number of $\GateIntSWAP$ operations is given by $2n$, which can be expressed as $2 \times \GateIntSWAP(i)$ for each layer, where $i$ ranges from $0$ to $n-1$. In more detail, this is equivalent to $2\sum^{n-1}_{i=1}2^{i-1}$, resulting in $2^{n} - 2$ pairs of $\GateCSWAP$S and $\GateSWAP$S. Additionally, there are two $\GateSWAP$s specifically for the root.

The next operation to consider is $\textsf{Routing}$, which is utilized in all three stages of the $(n,k)$-QRAM implementation. In the address setting and uncomputing phase, the $\textsf{Routing}$ operation is responsible for transferring each address data to layer $l$. This requires the execution of $2 \times \textsf{Routing}\left(i, i+1\right)$ operations, where $i$ ranges from $0$ to $l-1$. Each $\textsf{Routing}\left(i, i+1\right)$ operation involves $2^{i}$ pairs of operations. Consequently, the total number of $\GateCSWAP$s and $\GateSWAP$s required for $\textsf{Routing}$ in the address setting and uncomputing phase is $2\sum^{n-1}_{l=1}\sum^{l-1}_{i=0}2^{i} = 2\left(2^n - n - 1\right)$ pairs. In the data fetching phase, the number of pairs of two $\GateCSWAP$s needed for \textsf{Routing} depends on the relationship between $k$ and $n$.  If $k \geq n$, it needs $2i + \left(k-i\right)$ $\textsf{Routing}\left(n-i, n-i-1\right)$, which means $2^{n-i-1} \left[2i + \left(k-i\right)\right]$. In total, it requires $\sum^{n-1}_{i=1}2^{n-i-1} \left[2i + \left(k-i\right)\right]$. If $k < n$, the total sum of pairs would be $\sum^{k-1}_{i=1}2^{n-i-1} \left[2i + \left(k-i\right)\right] + 2k\sum^{n-1}_{j=k}2^{n-j-1}$. 

During the address setting and uncomputing phase, as well as the unidirectional $\textsf{Routing}$ operations in the data loading phase, all $\textsf{Routing}$ operations can be simplified using $\GateCiSCZ$s and $\GateiSCZ$s, thanks to the first extension facilitated by our strategy. This simplification is independent of the value of $k$. Consequently, the total number of $\GateCSWAP$s and $\GateSWAP$s saved can be calculated as $3 \times 2^n - 2n - 4 + \sum^{n-1}_{i=1}2^{n-i} \left[\min\left(i, k\right)\right]$. On the other hand, the remaining $\textsf{Routing}$ operations are all bidirectional, requiring $\sum^{n-1}_{i=1}2^{n-i-1} \left[k - \min\left(i, k\right)\right]$ pairs of two $\GateCSWAP$s.

\section{Detailed numerical simulations\label{sec: SWAP Network Numerical Simulation}}
    \paragraph{iSCZ pulse optimization}
        The waveform search is done using the differential evolution (DE) and the Nelder Mead (NM) optimization method together~\cite{ControlPulseForm_Machnes_2018, ToffoliOptimization_Zahedinejad_2015, NMOpt_Nelder_1965}. Due to the NM algorithm’s sensitivity to initial points, we first utilize the DE to get the proper starting point and then use NM to move forward. We use this framework to get three optimized pulses $\GateiSCZ$, $\textsf{fSim}$, and $\GateCZ$. The fidelity of quantum process tomography~\cite{ExperimentalFidelity_Molmer_2007} is calculated based on the Eq.~\eqref{eqn: experimental quantum operation fidelity}.
    \paragraph{Kraus operator simulation with $T_{1}$ and $T_{2e}$}
        The numerical experiment in Fig.~\ref{fig: Pulse and Transmon Model}(e) is on the performance using $\GateiSCZ$ to replace the $\GateSWAP$. We calculate the average fidelity~\cite{Nielson-QuditFidelity-2002} of the noisy channel considering the practical relaxation time $T_{1}$ and the echo dephasing time $T_{2e}$. We conduct numerical experiments with the parameters from 5 $\mu s$ to 50 $\mu s$, and the $T_{2e}$ is from $T_{1}$ to $2T_{1}$. In this simulation, the single-qubit gate time is set to zero. 
        
        The amplitude damping channel, denoted as $\mathcal{N}_{AD}$, is a mathematical model used to describe the decoherence processes that lead to energy dissipation in qubits. It represents how energy is lost from a qubit due to interaction with its environment. The action of the amplitude damping channel on an arbitrary quantum state, represented by the density matrix $\rho$, is given by

        \begin{equation}
            \mathcal{N}_{AD}(\rho) = E_0 \rho E_0^\dagger + E_1 \rho E_1^\dagger.
        \end{equation}
        
        In this expression, $E_0$ and $E_1$ are the Kraus operators that describe the different possible effects on the quantum state. The operators are defined as:
        
        \begin{equation}
            E_0 = \begin{pmatrix} 1 & 0 \\ 0 & \sqrt{1 - \gamma} \end{pmatrix}, \quad
            E_1 = \begin{pmatrix} 0 & \sqrt{\gamma} \\ 0 & 0 \end{pmatrix}.
        \end{equation}
        
        Here, $\gamma$ represents the damping probability, which describes the probability of energy loss from the qubit. The evolution of this damping probability over time is expressed as
        $\gamma(t) = 1 - e^{-t / T_1}$,
        where $T_1$ is known as the qubit relaxation time. It indicates how long the qubit remains in its excited state before decaying.
        
        In addition to the amplitude damping channel, another important type of noise affecting quantum systems is described by the dephasing channel, also called the phase damping channel, denoted by $\mathcal{N}_{PD}$. This channel models decoherence processes where the qubit's energy remains unchanged, but the phase information of the quantum state is disturbed. The dephasing channel acts on a qubit with a density matrix $\rho$ in a similar manner to the amplitude damping channel:
        
        \begin{equation}
            \mathcal{N}_{PD}(\rho) = E_0 \rho E_0^\dagger + E_1 \rho E_1^\dagger,
        \end{equation}
        
        where the Kraus operators are given by:
        
        \begin{equation}
            E_0 = \begin{pmatrix} 1 & 0 \\ 0 & \sqrt{1 - \lambda} \end{pmatrix}, \quad
            E_1 = \begin{pmatrix} 0 & 0 \\ 0 & \sqrt{\lambda} \end{pmatrix}.
        \end{equation}
        
        In this case, $\lambda$ represents the scattering probability, which is associated with phase disturbances in the quantum state without a loss of energy. The temporal evolution of the scattering probability \(\lambda(t)\) is given by $\lambda(t) = 1 - e^{\frac{t}{T_1} - \frac{2t}{T_{2e}}}$, where $T_1$ is the qubit relaxation time, and $T_{2e}$ is the qubit dephasing time.

        We first transform the noiseless and noisy operation(in $\chi$ matrix) into the Pauli transfer matrix~\cite{Superoperator_Wood_2015}. Based on the equation, we calculated the exact form of the average fidelity~\cite{Nielson-QuditFidelity-2002} under the frame of a superoperator~\cite{Coherence_Iverson_2020}. And the exact form is given by,  
        \begin{equation}
            r = \frac{1}{d (d + 1)} \text{Tr}(I_{d^2} - N),
        \end{equation}
        where $d = 2^n$ represents the dimension of the Hilbert space for a system with $n$ qubits, and $N$ is the Pauli transfer matrix that characterizes the quantum channel on the Pauli basis. 
        
    \paragraph{QRAM circuit depth and multi-qubit operation count}
        The QRAM circuit is implemented in QISKit~\cite{Qiskit} to count the number of $\GateCX$ gates. We utilize the default tranpilation method built in QISKit and try three different optimization levels. The depth and mulit-qubit operation count are presented in the Table.~\ref{tab: Gate Cost} is the lowest number get from the 10000 trials.
    
    \paragraph{Random SWAP network circuit simulation}
        We focus on a SWAP network, in which all qubits are arranged linearly, and $\GateSWAP$ gates are applied exclusively between physically adjacent qubits. We conducted a comparative analysis of circuit depth and fidelities of these two strategies for decomposing $\GateSWAP$ gates. 
        Given a size $N$, an initial list $[0, 1, \cdots, N-1]$ is generated as the original layout, and a random permutation of these $N$ integers is produced, resulting in a new list $[P(0), P(1), \cdots, P(N-1)]$. The value $P(i)$ at the $i$-th index of the new list represents the information previously stored in the $P(i)$-th qubit of the initial layout. Once the permutation is fixed, the routing path and the sequence of $\textsf{SWAP}$ gates are determined. For each size $N$, the experiment is repeated 100 times, resulting in 100 random permutations and corresponding layouts.
                
\begin{table}[h]
\caption{Gate Complexity of $\textsf{(n, k)-QRAM}$. Each entries of the table show the number of pairs of \textsf{SWAP} and \textsf{C-SWAP} that could be optimized, from which \textsf{Internal-SWAP} and Unidirectional $\textsf{Routing}$ are using Extension 1 while Entension 2 is reponsible for optimizing Bidirectional $\textsf{Routing}$.}
\vspace{0.1cm}
\centering
\begin{threeparttable}
\begin{tabular}{cc|cc|ccc}
\toprule[2pt]
\multicolumn{2}{c|}{} & \multicolumn{2}{c|}{Extension 1} & \multicolumn{1}{c}{Extension 2}          \\ 
\cmidrule[1pt]{3-5}
  \begin{tabular}[c]{@{}c@{}}\end{tabular} &
  \begin{tabular}[c]{@{}c@{}}\end{tabular} &
  \begin{tabular}[c]{@{}c@{}}\textsf{Internal-SWAP}\end{tabular} &
  \begin{tabular}[c]{@{}c@{}}Unidirectional \\ $\textsf{Routing}$\end{tabular} &
  \begin{tabular}[c]{@{}c@{}}Bidirectional \\ $\textsf{Routing}$\end{tabular} \\
\midrule[1pt]
\begin{tabular}[c]{@{}c@{}}Address Setting \\ \& Uncomputing\end{tabular}& &  $2^n -2$  & $2\left(2^n - n - 1\right)$ & -   \\ 
 \midrule[1pt]
 \begin{tabular}[c]{@{}c@{}}Data \\ Loading\end{tabular} & & -  & $\sum^{n-1}\limits_{i=1}2^{n-i} \left[\min\left(i, k\right)\right]$ &  $\sum^{n-1}\limits_{i=1} 2^{n-i-1} \left[k - \min\left(i, k\right)\right]$   \\ 
 \bottomrule[2pt]
\end{tabular}
\end{threeparttable}
\label{tab:experiment}
\end{table}

        

\end{document}